\documentclass[fleqn,usenatbib]{mnras}

\usepackage{newtxtext,newtxmath}
\usepackage[T1]{fontenc}
\usepackage{ae,aecompl}
\usepackage{pdflscape}
\usepackage{rotating}
\usepackage{longtable}
\usepackage{xtab,booktabs}
\usepackage{graphicx}
\usepackage{amsmath}	
\usepackage{multirow}

\usepackage{scalerel,tikz}
\usetikzlibrary{svg.path}
\definecolor{orcidlogocol}{HTML}{A6CE39}
\tikzset{orcidlogo/.pic={
 \fill[orcidlogocol] svg{M256,128c0,70.7-57.3,128-128,128C57.3,256,0,198.7,0,128C0,57.3,57.3,0,128,0C198.7,0,256,57.3,256,128z};
 \fill[white] svg{M86.3,186.2H70.9V79.1h15.4v48.4V186.2z}
 svg{M108.9,79.1h41.6c39.6,0,57,28.3,57,53.6c0,27.5-21.5,53.6-56.8,53.6h-41.8V79.1z M124.3,172.4h24.5c34.9,0,42.9-26.5,42.9-39.7c0-21.5-13.7-39.7-43.7-39.7h-23.7V172.4z}
 svg{M88.7,56.8c0,5.5-4.5,10.1-10.1,10.1c-5.6,0-10.1-4.6-10.1-10.1c0-5.6,4.5-10.1,10.1-10.1C84.2,46.7,88.7,51.3,88.7,56.8z};
}}
\newcommand\orcidicon[1]{\href{https://orcid.org/#1}{\mbox{\scalerel*{
\begin{tikzpicture}[yscale=-1,transform shape]
\pic{orcidlogo};
\end{tikzpicture}
}{|}}}}

\newcommand{\aref}[1]{\hyperref[#1]{Appendix~\ref{#1}}}

\newread\infile
\def\preparetable#1#2{\bgroup \openin\infile=#1
   \let\\=\relax \gdef\usetable{}\preparetableA #2,,}
\def\preparetableA #1,{\if,#1,\egroup \closein\infile \else \read\infile to\tmp
  \xdef\usetable{\usetable \tmp & #1 \\}\expandafter\preparetableA\fi}

\title[Kinematics and metallicity gradients at high-$z$]{The role of gas kinematics in setting metallicity gradients at high redshift}

\author[P. Sharda et al.]{Piyush Sharda$^{\orcidicon{0000-0003-3347-7094}\,1,2}$\thanks{piyush.sharda@anu.edu.au (PS)},
Emily Wisnioski$^{\orcidicon{0000-0003-1657-7878}\,1,2}$\thanks{emily.wisnioski@anu.edu.au (EW)},
Mark R. Krumholz$^{\orcidicon{0000-0003-3893-854X}\,1,2}$\thanks{mark.krumholz@anu.edu.au (MRK)}, and
Christoph Federrath$^{\orcidicon{0000-0002-0706-2306}\,1,2}$\\
$^{1}$Research School of Astronomy and Astrophysics, Australian National University, Canberra, ACT 2611, Australia\\
$^{2}$Australian Research Council Centre of Excellence for All Sky Astrophysics in 3 Dimensions (ASTRO 3D), Australia\\
}

\date{Accepted 2021 June 25. Received 2021 June 02; in original form 2021 January 26}

\pubyear{2021}

\begin{document}
\label{firstpage}
\pagerange{\pageref{firstpage}--\pageref{lastpage}}
\maketitle

\begin{abstract}
In this work, we explore the diversity of ionised gas kinematics (rotational velocity $v_{\phi}$ and velocity dispersion $\sigma_{\rm{g}}$) and gas-phase metallicity gradients at $0.1 \leq z \leq 2.5$ using a compiled data set of 74 galaxies resolved with ground-based integral field spectroscopy. We find that galaxies with the highest and the lowest $\sigma_{\rm{g}}$ have preferentially flat metallicity gradients, whereas those with intermediate values of $\sigma_{\rm{g}}$ show a large scatter in the metallicity gradients. Additionally, steep negative gradients appear almost only in rotation-dominated galaxies ($v_{\phi}/\sigma_{\rm{g}} > 1$), whereas most dispersion-dominated galaxies show flat gradients. We use our recently developed analytic model of metallicity gradients to provide a physical explanation for the shape and scatter of these observed trends. In the case of high $\sigma_{\rm{g}}$, the inward radial advection of gas dominates over metal production and causes efficient metal mixing, thus giving rise to flat gradients. For low $\sigma_{\rm{g}}$, it is the cosmic accretion of metal-poor gas diluting the metallicity that gives rise to flat gradients. Finally, the reason for intermediate $\sigma_{\rm{g}}$ showing the steepest negative gradients is that both inward radial advection and cosmic accretion are weak as compared to metal production, which leads to the creation of steeper gradients. The larger scatter at intermediate $\sigma_{\rm{g}}$ may be due in part to preferential ejection of metals in galactic winds, which can decrease the strength of the production term. Our analysis shows how gas kinematics play a critical role in setting metallicity gradients in high-redshift galaxies.
\end{abstract}
    


\section{Introduction}
\label{s:intro}
Understanding the distribution of metals in galaxies is crucial to learn about galaxy formation and evolution. It is now well known that metals in both the gas and stars show a negative, radial gradient across the discs of most galaxies. Since the discovery of metallicity gradients in galactic discs \citep{1942ApJ....95...52A,1971ApJ...168..327S,1983MNRAS.204...53S}, several attempts have been made to put the measurements in context of galaxy evolution theory, as well as understand the physics driving the magnitude of the gradient by exploring trends with galaxy properties, such as mass, star formation rate (SFR), star formation efficiency, specific SFR, radial inflows, cosmic infall, etc. \citep[see recent reviews by][]{2019A&ARv..27....3M,2019ARA&A..57..511K,2020arXiv200900424S,2020ARA&A..58...99S,2020ARA&A..58..661F}. With the advent of large resolved spectroscopic surveys using integral field unit (IFU) spectroscopy we are now able to explore the relationship between metallicity gradients and galaxy kinematics (\textit{i.e.,} the rotational velocity $v_{\phi}$ and the velocity dispersion $\sigma_{\rm{g}}$). There are several reasons why we would expect such a correlation to exist. For example, turbulent mixing and transport, processes whose rates are expected to scale with $\sigma_{\rm{g}}$, should be important processes that influence metallicity gradients \citep[e.g.,][]{2012ApJ...758...48Y,2014MNRAS.438.1552F,2015MNRAS.449.2588P,2018MNRAS.481.5000A,2018MNRAS.475.2236K,2020MNRAS.499..193K,2021MNRAS.504.5496L}. Similarly, rates of cosmic infall that can dilute both the overall metallicity and its gradients should correlate strongly with halo mass, which is closely linked to $v_{\phi}$ \citep[e.g.,][]{1977A&A....54..661T,2000ApJ...533L..99M,2001ApJ...550..212B,2011MNRAS.417.2982F}. The production of metals is dictated by star formation in galaxies, and star formation feedback also impacts galaxy kinematics \citep[e.g.,][]{2011ApJ...731...41O,2013MNRAS.433.1970F,2014MNRAS.438.1552F,2015ApJ...815...67K,2016ApJ...827...28G,2018MNRAS.477.2716K,2021MNRAS.500.3394F}. The amount of metals lost in outflows is also expected to scale inversely with $v_{\phi}$ \citep{2002ApJ...581.1019G}. Thus, there are several links between metallicity gradients and galaxy kinematics, and it is clear that these links likely generate a rather complex relationship between each other as well as other relevant mechanisms.

\begin{table*}
\centering
\caption{Summary of the data adopted from different sources in the literature. Columns~$1-3$ list the different samples, instruments used to measure emission lines and the number of galaxies $N$ that we use from each sample, respectively. Columns~$4-5$ list the range in redshift and stellar mass of the observed galaxies. Column~$6$ lists the spectral resolution for each instrument, and columns~$7$ and $8$ list the PSF FWHM in arcsec and kpc, respectively. Finally, column~$9$ lists the references for each sample: (a.) \protect\cite{2012A&A...539A..92E}, (b.) \protect\cite{2012A&A...539A..93Q}, (c.) \protect\cite{2013ApJ...779..139S}, (d.) \protect\cite{2014MNRAS.443.2695S}, (e.) \protect\cite{2012MNRAS.426..935S}, (f.) \protect\cite{2018MNRAS.478.4293C}, (g.) \protect\cite{2018ApJS..238...21F}.}
\begin{tabular}{|l|c|c|c|c|c|c|c|r}
\hline
Sample & Instrument & $N$ & $z$ & $\log_{10} M_{\star}/\mathrm{M_{\odot}}$ & $R$ & PSF FWHM (\arcsec) & PSF FWHM (kpc) & Ref.\\
(1) & (2) & (3) & (4) & (5) & (6) & (7) & (8) & (9)\\
\hline
\textit{MASSIV} & SINFONI & 19 & $0.9-1.6$ & $9.4-11.0$ & $2000-2640$ & $0.3-1.0$ & $2-7$ & a, b \\
\textit{HiZELS} & KMOS & 9 & $\approx 0.81$ & $9.8-10.7$ & $\approx 3400$ & $\sim 0.8$ & $\sim 6$ & c, d \\
\textit{SHiZELS} & SINFONI & 6 & $0.8-2.2$ & $9.4-11.0$ & $\approx 4500$ & $\sim 0.1$ & $0.7-0.8$ & e \\
\textit{MUSE-WIDE} & MUSE & 23 & $0.1-0.8$ & $8.3-10.6$ & $1650-3800$ & $0.6-0.7$ & $1-5$ & f \\
\textit{SINS\,/\,zC-SINF} & SINFONI & 17 & $1.4-2.4$ & $10.1-11.5$ & $2730-5090$ & $0.1-0.3$ & $\sim 0.8$ & g \\
\hline
\end{tabular}
\label{tab:tab1}
\end{table*}

This connection is perhaps most readily explored at high redshift ($z \leq 2.5$), when galaxies show a more diverse range of metallicity gradients and kinematics than are found in the local Universe \citep{2019A&ARv..27....3M,2020ARA&A..58..661F,2020ARA&A..58..157T}. The last decade has seen immense progress in these areas, thanks to IFU spectroscopy instruments like MUSE \citep[Multi Unit Spectroscopic Explorer,][]{2010SPIE.7735E..08B}, KMOS \citep[K-band Multi Object Spectrograph,][]{2004SPIE.5492.1179S}, SINFONI \citep[Spectrograph for INtegral Field Observations in the Near Infrared,][]{2003SPIE.4841.1548E,2004SPIE.5490..130B}, FLAMES \citep[Fibre Large Array Multi Element Spectrograph,][]{2002Msngr.110....1P}, GMOS \citep[Gemini Multi Object Spectrograph,][]{1997SPIE.2871.1099D}, NIFS \citep[Gemini Near-infrared Integral Field Spectrograph,][]{2003SPIE.4841.1581M}, and OSIRIS \citep[OH-Suppressing InfraRed Imaging Spectrograph,][]{2006SPIE.6269E..1AL}.

Studies using these instruments have revealed that, while high-$z$ galaxies show a diverse range of metallicity gradients, the average evolution of these gradients is rather shallow, almost non-existent \citep[Figure~8]{2020MNRAS.492..821C}. On the other hand, there is ample evidence for redshift evolution of galaxy kinematics. In particular, $\sigma_{\rm{g}}$ evolves with $z$ implying that high-$z$ discs are thicker and more turbulent \citep{2012ApJ...758..106K,2015ApJ...799..209W,2019ApJ...886..124W,2017ApJ...843...46S,2019ApJ...880...48U}. The mass-averaged rotational velocities are also expected to evolve with time (e.g., \citealt{2011MNRAS.410.1660D,2016MNRAS.460..103T,2017ApJ...839...57S,2017ApJ...842..121U,2017MNRAS.466.4780M,2020arXiv201108866G}; see, however, \citealt{2019MNRAS.482.2166T}). However, links between kinematics and metallicity gradients at high redshift have been investigated by observations only in a handful of studies \citep{2012A&A...539A..93Q,2021MNRAS.500.4229G}, most of which were limited to gravitationally-lensed samples \citep{2011ApJ...732L..14Y,2013ApJ...765...48J,2016ApJ...820...84L}, yielding no clear connections between the two. Some simulations have also started to explore joint evolution of metallicity gradients and kinematics \citep{2017MNRAS.466.4780M,2020arXiv200710993H}, but at present theoretical work is limited to empirical examination of simulations results. No models proposed to date have quantitatively discussed the observed correlations between metallicity gradients and gas kinematics.

In a companion paper \citep{2020aMNRAS.xxx..xxxS}, we presented a new model for the physics of gas phase metallicity gradients from first principles. We showed that our model successfully reproduces several trends of metallicity gradients with galaxy properties, for example, the observed cosmic evolution of metallicity gradients \citep{2020aMNRAS.xxx..xxxS} and the mass-metallicity gradient relation (MZGR, \citealt{2020bMNRAS.xxx..xxxS}). The goal of this paper is to apply the model to existing observations of high-redshift galaxies to investigate the relationship between metallicity gradients and gas kinematics.

This paper is organized as follows: \autoref{s:data} describes the data on metallicity gradients and galaxy kinematics that we compile from observations, \autoref{s:results} presents the resulting trends we find in the data, \autoref{s:compare_theory} presents a discussion on the comparison of the observational data with our theoretical model, and \autoref{s:conclusion} lists our conclusions. For this work, we use the $\Lambda$CDM cosmology with $H_0 = 71\,\mathrm{km\,s^{-1}\,Mpc^{-1}},\,\Omega_{\mathrm{m}} = 0.27$ and $\Omega_\Lambda = 0.73$ \citep{2003MNRAS.339..289S}. Further, we express $\mathcal{Z} = Z/\rm{Z_{\odot}}$, where $\rm{Z_{\odot}} = 0.0134$ \citep{2009ARA&A..47..481A}, and we use the \citet{2003PASP..115..763C} initial mass function (IMF).

\begin{figure*}
\includegraphics[width=1.0\linewidth]{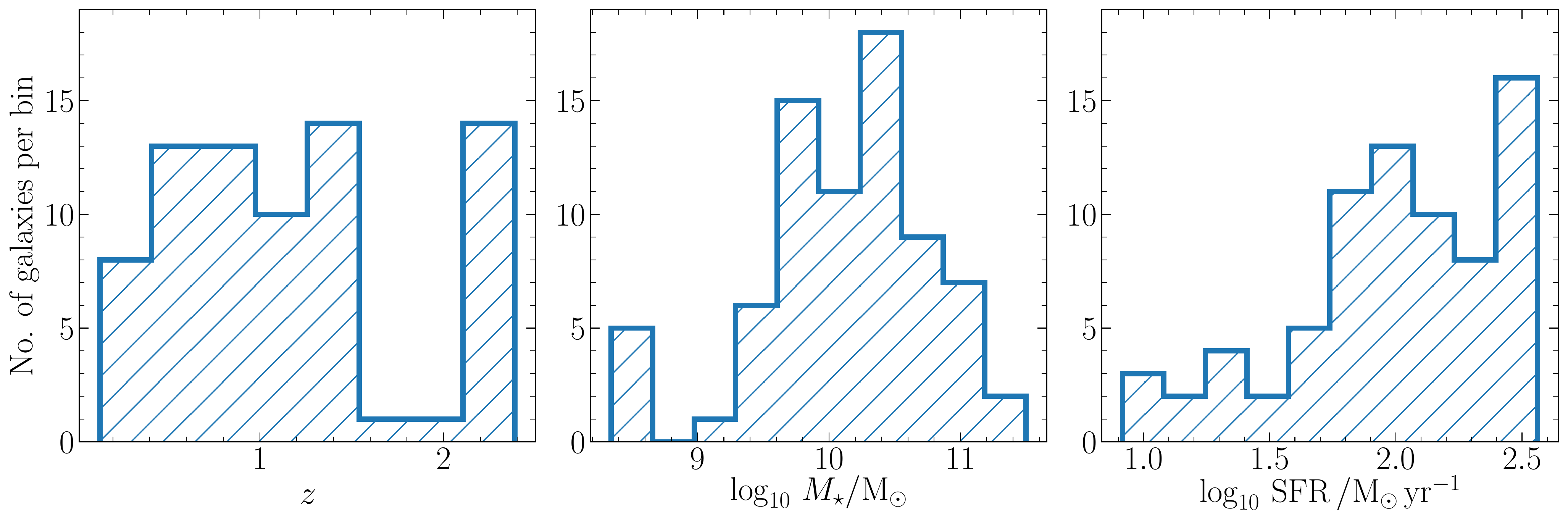}
\caption{Distribution of galaxies at different redshifts (left), stellar mass (middle) and star formation rate (right) in the compiled sample used in this work.}
\label{fig:pdfs1}
\end{figure*}

\section{Compiled Data and Analysis}
\label{s:data}
We compile a sample of 74~non-lensed high-$z$ ($0.1 \leq z \leq 2.5$) galaxies from the literature, studied with ground-based IFU instruments suitable for the measurement of metallicity gradients and gas kinematics. We only work with non-lensed galaxies because it is not yet clear if lens reconstructions accurately reproduce metallicity maps \citep[Section~6.7]{2019A&ARv..27....3M}. However, we note that there is a similiar diversity of gradients from lensed galaxies \citep{2016ApJ...827...74W,2016ApJ...820...84L}, and including them does not change our results. We describe each of the samples we use in \autoref{s:samples}, and provide a summary in \autoref{tab:tab1}. Our database is inhomogeneous, because the sources we draw from have different sample selections, varying resolution, and use different techniques to obtain the metallicity gradients and kinematics. To alleviate some of the inhomogeneity, we reanalyse the kinematics using the same method for the full database, a process that we describe in \autoref{s:kinematicreanalysis}. Additionally, where possible, we use a common metallicity diagnostic and calibration to estimate metallicity gradients. We list the database along with the reanalysed kinematics for all 74~galaxies in \aref{s:app_data}.  

We acknowledge that there are many challenges associated with measuring metallicity gradients and kinematics in IFU observations, particularly at high-$z$. Metallicity measurements in \ion{H}{ii} regions rely on accurately modeling the \ion{H}{ii}-region physics, and systematic variations in the physical parameters with redshift, if any \citep[e.g.,][]{2013ApJ...774..100K,2014ApJ...787..120S,2018ApJ...868..117S,2020arXiv201210445D}, emission line diagnostics and calibrations \citep[e.g.,][]{2008ApJ...681.1183K,10.1093/mnras/stab205}, spatial and spectral resolution \citep[e.g.,][]{2013ApJ...767..106Y,2014A&A...561A.129M}, and contamination from shocks and active galactic nuclei (AGN, \citealt{2013ApJ...774L..10K,2014ApJ...781...21N}). Similarly, kinematic measurements rely on model assumptions, source blending, beam smearing, and spectral resolution limits \citep[e.g.,][]{2011ApJ...741...69D,2015MNRAS.451.3021D,2016ApJ...826..214B,2018ApJ...855...97W}. Thus, systematic errors originating from these physical and observational effects should be kept in mind in the context of our work. We note that some of the kinematic uncertainties are not captured in the quoted errors, which only account for uncertainties in the beam smearing, inclination, and instrumental resolution corrections.

\subsection{Samples}
\label{s:samples}

\begin{enumerate}
    \item \textit{MASSIV.} We use data from the Mass Assembly Survey with SINFONI in VIMOS VLT Deep Survey \citep[MASSIV,][]{2012A&A...539A..91C} of star-forming galaxies between $1 < z < 2$. The kinematics and the metallicity gradients from this survey are described in \cite{2012A&A...539A..92E} and \cite{2012A&A...539A..93Q}, respectively. The authors report on metallicity gradients using the [\ion{N}{ii}]/H$\alpha$ ratio following the \cite{2009MNRAS.398..949P} calibration. In order to be consistent with the other samples described below, we reverse the \cite{2009MNRAS.398..949P} calibration to obtain the [\ion{N}{ii}]/H$\alpha$ flux at different locations in the galactic disc, and use the flux to find the metallicities using the \cite{2004MNRAS.348L..59P} calibration. We then use metallicities based on the \cite{2004MNRAS.348L..59P} calibration to measure the metallicity gradients. We reanalyse the kinematics for this sample following the procedure described below in \autoref{s:kinematicreanalysis}.
    \item \textit{HiZELS.} This sample consists of galaxies at $z \sim 0.8$ that were observed through KMOS as part of the High-$z$ Emission Line Survey \citep[HiZELS,][]{2009MNRAS.398...75S,2013MNRAS.428.1128S}. The kinematics for these galaxies are reported in \cite{2013ApJ...779..139S} and the metallicity gradients in \cite{2014MNRAS.443.2695S}. The metallicity gradients are measured with the [\ion{N}{ii}]/H$\alpha$ ratio using the \cite{2004MNRAS.348L..59P} calibration.
    \item \textit{SHiZELS.} In addition to the \textit{HiZELS} survey above, we also use observations from the SINFONI-HiZELS survey \citep[SHiZELS,][]{2012MNRAS.426..935S} that report on metallicity gradients and kinematics of 8 galaxies in the redshift range $0.8-2.2$. The gradients are measured with the [\ion{N}{ii}]/H$\alpha$ ratio using the \cite{2004MNRAS.348L..59P} calibration.   
    \item \textit{MUSE-WIDE.} We take measurements of metallicity gradients carried out by \cite{2018MNRAS.478.4293C} for galaxies at low redshift ($0.08 < z< 0.84$) using MUSE. The authors use a forward-modeling Bayesian approach to estimate the metallicity gradients \citep{2017MNRAS.468.2140C} from nebular emission lines (for $z \leq 0.4$, H$\beta$, \ion{O}{iii}, H$\alpha$, and \ion{S}{ii}, and for $z > 0.4$, \ion{O}{ii}, H$\gamma$, H$\beta$, and \ion{O}{iii}). The kinematics for these galaxies are not available in the literature, so we obtain them by fitting publicly-available data \citep{2017A&A...606A..12H,2019A&A...624A.141U} using the emission line fitting code LZIFU \citep{2016Ap&SS.361..280H}. We describe this in detail in \autoref{s:kinematicreanalysis}. We obtain the half-light radii for these galaxies from The Cosmic Assembly Near-infrared Deep Extragalactic Legacy Survey \citep[CANDELS,][]{2012ApJS..203...24V} and from 3D-HST photometry \citep{2014ApJS..214...24S}.
    \item \textit{SINS\,/\,zC-SINF.} \cite{2018ApJS..238...21F} report SINFONI observations of metallicity gradients and kinematics in galaxies at $z \sim 1.5-2.2$ from the \textit{SINS\,/\,zC-SINF} survey, where the authors use the [\ion{N}{ii}]/H$\alpha$ ratio to quantify the metallicity gradient. In order to homogenise their sample with other samples above, we use their recommended conversion factor to scale the gradients to the calibration given by \cite{2004MNRAS.348L..59P}. The reported kinematics for this sample are already corrected for instrumental and beam smearing effects using the approach from \cite{2016ApJ...826..214B} which we utilise for the other samples in \autoref{s:kinematicreanalysis}. 
\end{enumerate}

All the above surveys also include information on the stellar mass $M_{\star}$ (scaled to the Chabrier IMF where required) and the dust-corrected SFR from H$\alpha$, except for the \textit{SINS\,/\,zC-SINF} survey. To obtain dust-corrected SFR estimates for \textit{SINS\,/\,zC-SINF}, we use the integrated H$\alpha$ fluxes reported by the authors, and scale them to find the dust-corrected H$\alpha$ luminosity following \cite{2001PASP..113.1449C}, and convert it to SFR based on the \citet{2003PASP..115..763C} IMF following \cite{2012ARA&A..50..531K}.

\autoref{fig:pdfs1} shows the distributions of redshift, stellar mass and SFR of galaxies in our compiled sample from the above surveys. The distribution in redshift is quite uniform, except around $z \approx 1.7$, where there is no available data due to atmospheric absorption. This implies that the data we use are not biased towards a particular redshift. It is also clear from \autoref{fig:pdfs1} that the observations consist primarily of more massive ($M_{\star} > 10^{10}\,\rm{M_{\odot}}$) galaxies; the few low-mass galaxies ($M_{\star} \leq 10^{9.5}\,\rm{M_{\odot}}$) that we are able to study belong to the \textit{MUSE-WIDE} sample. The overall sample is somewhat biased to high star formation rates: 20 per cent of the galaxies in the compiled sample have SFRs more than $3\times$ the main sequence SFR for their mass and redshift \citep{2012ApJ...754L..29W}. This bias is not surprising, given that large H$\alpha$ fluxes (corresponding to large SFRs) are typically necessary for spatially-resolved measurements at high redshift. However, we emphasise that the sample is not dominated by merging or interacting galaxies: based on the classifications provided by the source papers from which we draw the sample, less than $9$ per cent of the galaxies are mergers or interactions. This means that our sample is not significantly affected by the flattening of gradients that typically occurs when galaxies merge \citep[e.g.,][]{2010ApJ...723.1255R,2012ApJ...753....5R,2014MNRAS.438.1894T,2017MNRAS.472.4404S}.

\begin{figure*}
\includegraphics[width=0.285\linewidth,angle=90]{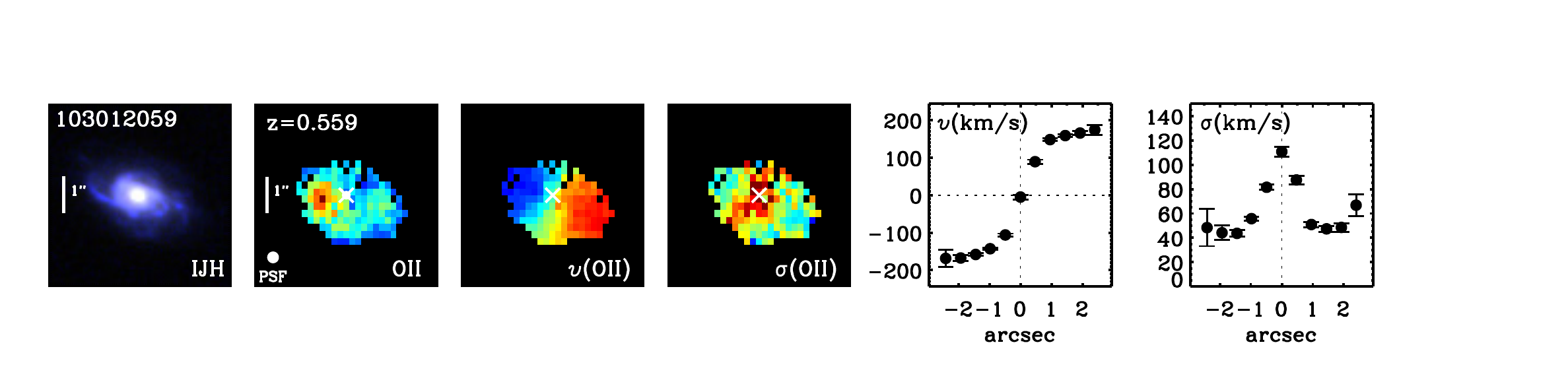}
\caption{\textit{Left to right} $-$ observed-frame IJH color composite image from CANDELS HST imaging \protect\citep{2011ApJS..197...35G,2011ApJS..197...36K}, nebular line flux with the strongest emission (\ion{O}{ii} in this case), rotational velocity $v$, velocity dispersion $\sigma$, as well as the 1D radial curves of $v$ and $\sigma$ derived from kinematic extractions for the galaxy G103012059 from the \textit{MUSE-WIDE} sample.}
\label{fig:app_kin1}
\end{figure*}

\begin{figure*}
\includegraphics[width=1.0\linewidth]{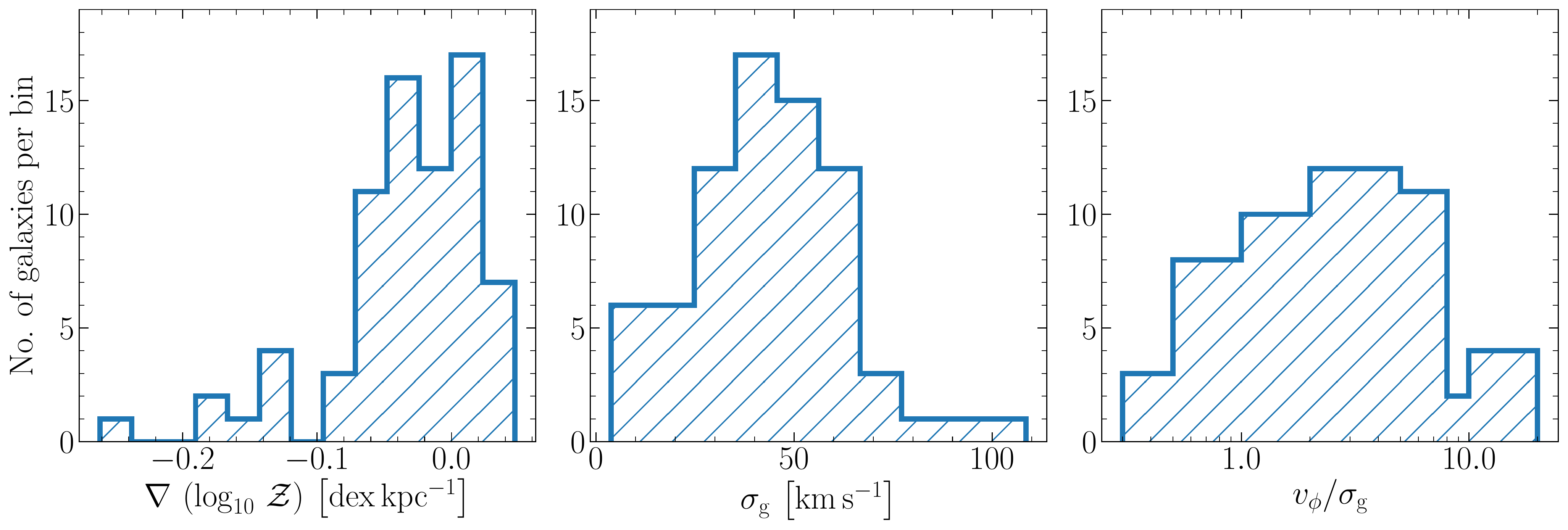}
\caption{Same as \autoref{fig:pdfs1}, but for the measured metallicity gradients (left), and reanalysed kinematics -- velocity dispersion $\sigma_{\rm{g}}$ (middle) and the ratio of rotational velocity to velocity dispersion $v_{\phi}/\sigma_{\rm{g}}$ (right).}
\label{fig:pdfs2}
\end{figure*}

\subsection{Kinematics}
\label{s:kinematicreanalysis}
To obtain the global kinematics for each galaxy, rotational velocity $v_{\phi}$ and characteristic velocity dispersion $\sigma_{\rm{g}}$, we use 1D velocity and dispersion curves extracted along the kinematic major axis following \cite{2019ApJ...886..124W}. Briefly, we measure the observed velocity by calculating the average of the absolute value of the minimum and maximum velocity measured along the kinematic axis and correcting for inclination. The measured velocity dispersion is calculated by taking the weighted mean of the outer data points of the 1D velocity dispersion profile. We adopt this non-parametric analysis to enable the use of galaxies with a variety of kinematic classifications. By not limiting the sample to the highest signal-to-noise disc galaxies we can investigate the metallicity gradients of galaxies with kinematic perturbations.

One dimensional kinematic extractions are directly provided for both the \textit{HiZELS} and \textit{SHiZELS} samples. For the \textit{MUSE-WIDE} and \textit{MASSIVE} samples the 1D kinematic profiles need to be measured. We use the datacubes for \textit{MASSIV} (B.~Epinat, private communication) and \textit{MUSE-WIDE} \citep{2017A&A...606A..12H,2019A&A...624A.141U} samples to derive these. We fit the data with the emission line diagnostic package LZIFU \citep{2016Ap&SS.361..280H}. LZIFU runs spectral decomposition on IFU datacubes to produce 2D emission line and kinematic maps based on the Levenberg-Marquardt least squares method.

We first produce emission line and kinematic maps for the entire galaxy by passing the complete datacube to LZIFU. We supply an external continuum map to LZIFU that we create by finding the median flux for every spatial pixel (spaxel; e.g., \textit{MUSE-WIDE}) or where the signal-to-noise of the continuum is negligible we simply supply a null external continuum map for the galaxies (e.g., \textit{MASSIV}). We use the resulting flux and moment-1 maps from the fit to locate the galaxy centre and the kinematic major axis, respectively. Once the kinematic major axis and the galaxy centre are determined, we create apertures with the size of the full width at half maximum (FWHM) of the PSF across the major axis. We sum the flux in each spaxel within these apertures along the major axis. This gives a spatially-summed spectrum for every aperture, thus increasing the signal to noise ratio. We then fit the aperture spectra with LZIFU, which returns a single value of $v_{\phi}(r)$ and $\sigma_{\rm{g}}(r)$ for every aperture that we use to create 1D radial curves.

After we derive the global velocities and dispersions from the 1D radial curves for all galaxies in the \textit{MASSIV}, \textit{HiZELS}, \textit{SHiZELS} and \textit{MUSE-WIDE} samples, we apply inclination, instrumental resolution, and beam smearing corrections on them. To correct for inclination, we simply divide the observed velocities by $\mathrm{sin}(i)$, where $i$ is the inclination angle. We use the inclinations reported in the source papers for this purpose. Following \cite{2015ApJ...799..209W}, we add a 30 per cent uncertainty in quadrature to $\sigma_{\rm{g}}$ if it is comparable to the instrumental resolution; if $\sigma_{\rm{g}}$ is less than the instrumental resolution, we add a 60 per cent uncertainty in quadrature. To correct for beam smearing, we follow \citet[Appendix~A2]{2016ApJ...826..214B}, as done by \citet[N. Förster-Schreiber, private communication]{2018ApJS..238...21F}. We note that this model makes the assumption that the galaxy kinematics are well described by a simple disc model. This may not apply to all galaxies in our sample, thus providing an over-correction for the beam in certain cases. The model assumes a Gaussian PSF and returns the beam-smearing correction factor based on the ratio of the stellar effective half-light radius to the beam effective half-light radius ($r_{\mathrm{e}}/r_{\mathrm{e,b}}$)\footnote{The measurements of the half-light radius for the different samples are based on broadband photometry using different bands, however it has a negligible effect on the beam smearing correction factor \citep{2016ApJ...828...27N}.}, and the ratio of the radius where the rotational velocity is calculated to the galactic effective half-light radius ($r_{\mathrm{vel}} / r_{\mathrm{e}}$). While it is straightforward to use these ratios to calculate the beam-smearing correction factor for $v_{\phi}$, those for $\sigma_{\rm{g}}$ also depend on the mass, inclination, and redshift of the source. We incorporate a 40 per cent error in $\sigma_{\rm{g}}$ to account for uncertainties in the beam smearing correction model \citep[Section~3.3]{2018ApJ...855...97W}, however it is possible that we may overestimate or underestimate the correction factor in certain cases. We present the resulting kinematics for all galaxies in \aref{s:app_data}, and illustrate the reanalysis procedure through a representative galaxy G103012059 (from the \textit{MUSE-WIDE} sample) in \autoref{fig:app_kin1}.

\subsection{Final sample}
We do not include all the galaxies that are available in the compiled surveys. We only select galaxies where the ratio of the radius at which $v_{\phi}$ is measured ($r_{\rm{vel}}$) to the half-light radius, $r_{\rm{e}}$, is greater than unity, as the beam-smearing correction model for $v_{\phi}$  requires $r_{\mathrm{vel}} > r_{\mathrm{e}}$. We also remove galaxies that only contain 3 or fewer resolution elements in our kinematic reanalysis, because we cannot derive a reasonable value for $v_{\phi}$ and $\sigma_{\rm{g}}$ in such cases. Further, we note that all the samples above exclude galaxies that contain contamination from active galactic nuclei (AGN), as diagnosed using the criteria described in \cite{2001ApJS..132...37K,2006MNRAS.372..961K} based on the \citet[BPT]{1981PASP...93....5B} diagram. The exception to this statement is the \textit{SINS\,/\,zC-SINF} sample, where the corresponding authors explicitly remove the contamination in gradients due to AGN for some of their galaxies. Our final sample consists of 74 galaxies with measured metallicity gradients and gas kinematics.

\begin{figure}
\includegraphics[width=1.0\columnwidth]{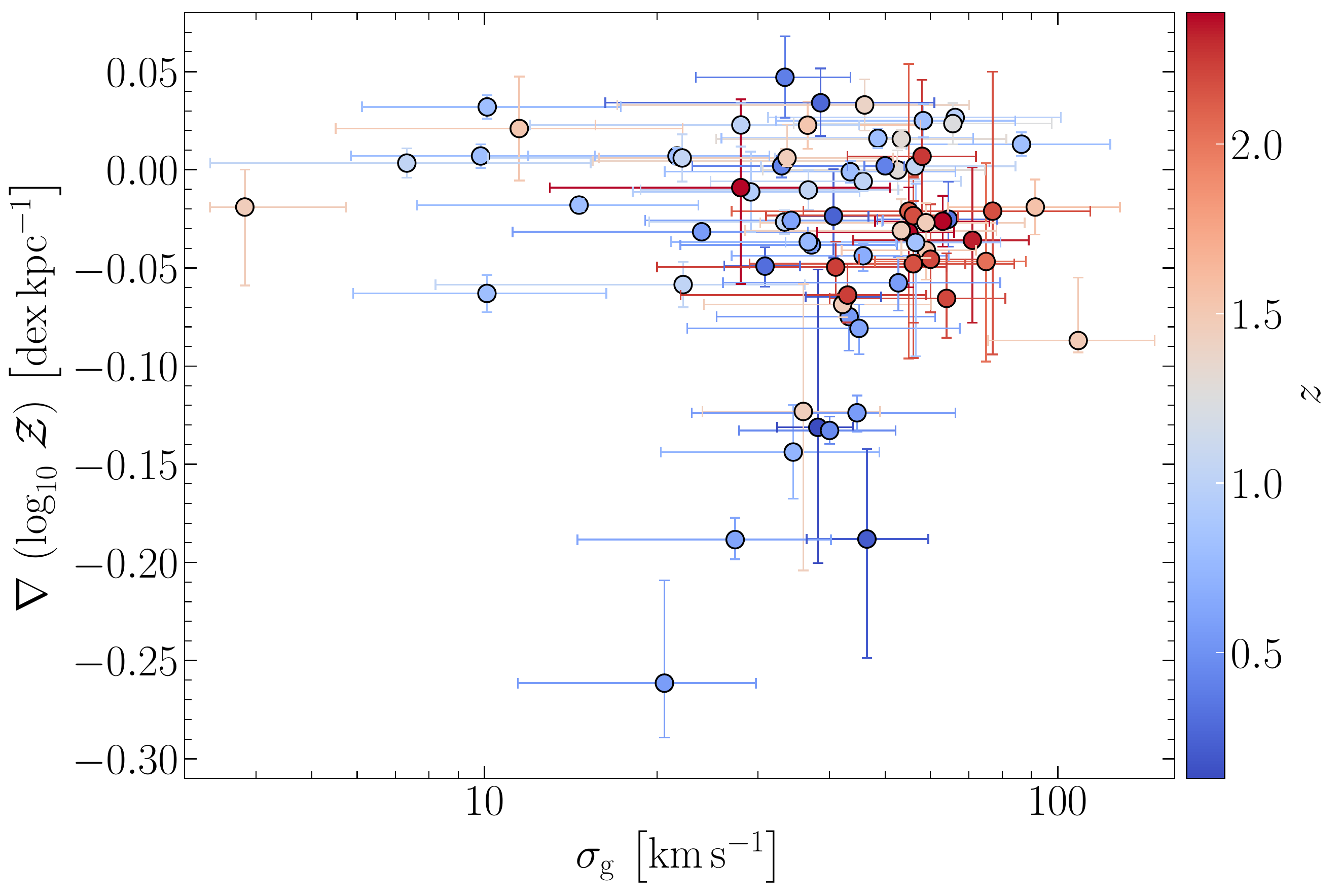}
\caption{Metallicity gradients in the compiled sample of high-redshift galaxies plotted as a function of velocity dispersion $\sigma_{\rm{g}}$, color-coded by redshift. We use the same method to derive the kinematics of all galaxies in our sample (see \autoref{s:kinematicreanalysis} for details). The quoted errorbars include uncertainties due to inclination, instrumental resolution, and beam smearing \protect\citep{2015ApJ...799..209W,2018ApJ...855...97W,2016ApJ...826..214B}.}
\label{fig:grad_sigma}
\end{figure}

\begin{figure}
\includegraphics[width=1.0\columnwidth]{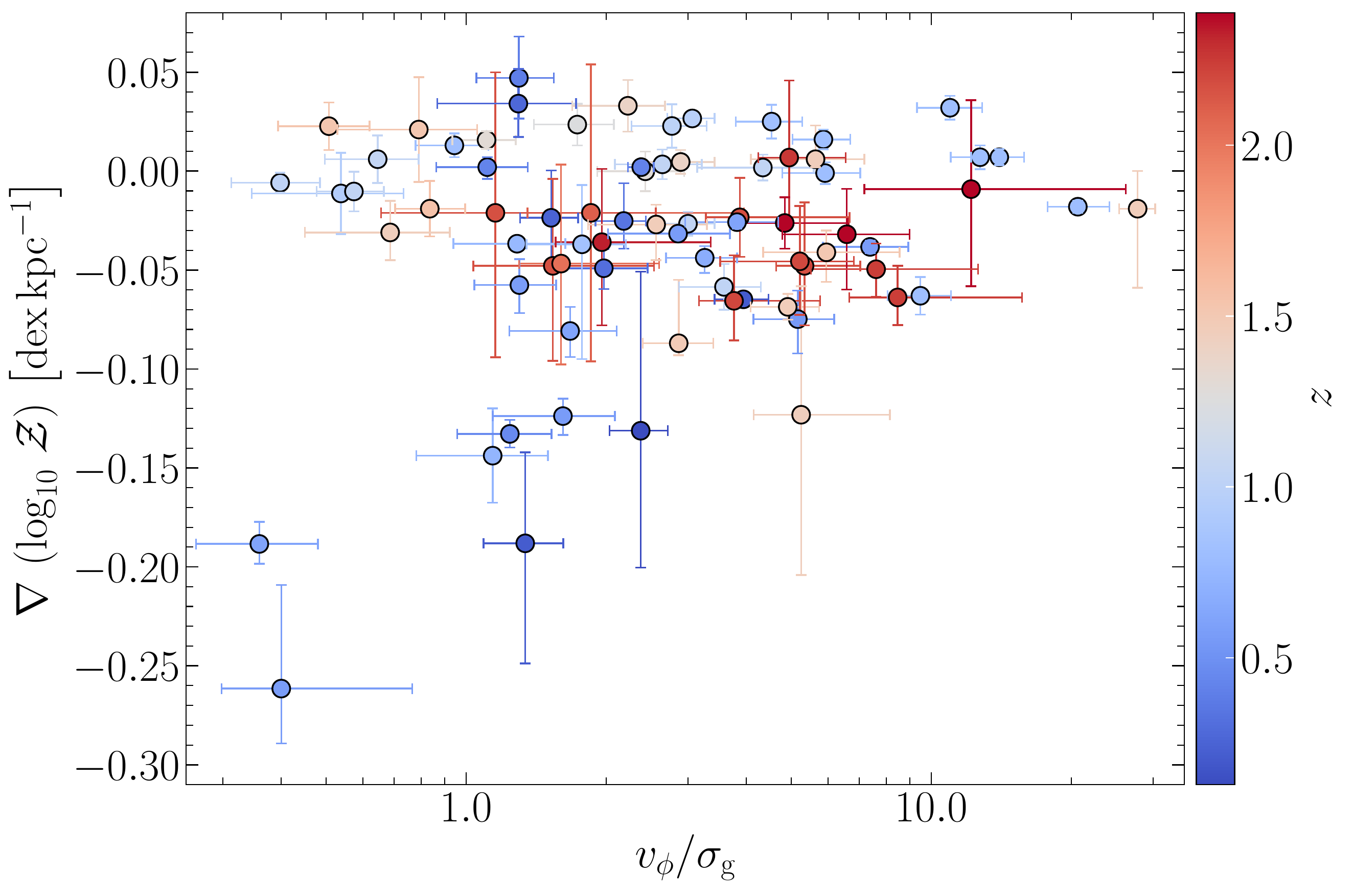}
\caption{Same data as \autoref{fig:grad_sigma}, but plotted against the ratio of rotational velocity to velocity dispersion, $v_{\phi}/\sigma_{\rm{g}}$. Galaxies with $v_{\phi}/\sigma_{\rm{g}} \geq 1$ are classified as rotation-dominated (and typically have a well-defined disc) whereas others are classified as dispersion-dominated (and typically have irregular structures).}
\label{fig:grad_vsigma}
\end{figure}

\section{Results}
\label{s:results}
\autoref{fig:pdfs2} shows the distributions of the metallicity gradients, and the resulting homogenised kinematics for all galaxies in the compiled data set. Our compilation recovers the diversity of gradients seen in the literature (by design) as well as the diversity of kinematics. This diversity is crucial for us to explore the correlations between galaxy kinematics and metallicity gradients, and is the primary driver of our work. We do not preferentially select disc galaxies however samples are likely to consist of primarily disc dominated galaxies due to the high fraction of discs at these epochs \citep{2015ApJ...799..209W,2019ApJ...886..124W,2016MNRAS.457.1888S,2017ApJ...843...46S}. Another indicator of the galaxy kinematics is the ratio of the rotational velocity to the velocity dispersion, $v_{\phi}/\sigma_{\rm{g}}$ \citep[e.g.,][]{2009ApJ...706.1364F,2018ApJS..238...21F,2010ApJ...725.2324B,2014ApJ...790...89K,2015ApJ...799..209W,2018ApJ...855...97W,2019ApJ...886..124W,2015AJ....149..107J,2019ApJ...874...59S}, which is used to determine the rotational support of galaxies. We find from \autoref{fig:pdfs2} that while most galaxies in the sample are rotation-dominated (\textit{i.e.,} $v_{\phi}/\sigma_{\rm{g}} \gtrsim 1$), around 18 per cent of the galaxies are dispersion-dominated ($v_{\phi}/\sigma_{\rm{g}} \lesssim 1$).

\autoref{fig:grad_sigma} shows the measured metallicity gradients as a function of the reanalysed velocity dispersion $\sigma_{\rm{g}}$, color-coded by redshift. We consider metallicity gradients as shallow or flat if the absolute strength of the gradient is less than $0.05\,\rm{dex\,kpc^{-1}}$. We observe that all galaxies in the sample with both high $\sigma_{\rm{g}} (\gtrsim 60\,\rm{km\,s^{-1}}$) and low $\sigma_{\rm{g}} (\lesssim 20\,\rm{km\,s^{-1}}$) show shallow or flat metallicity gradients, though we caution that the small velocity dispersions suffer significant uncertainties, as discussed in \autoref{s:data}. By contrast, galaxies with intermediate $\sigma_{\rm{g}}$ ($20-50\,\rm{km\,s^{-1}}$) show both the steepest gradients ($\sim -0.25\,\rm{dex\,kpc^{-1}}$) and the largest scatter ($\sim 0.1$) in gradients. We find only one galaxy with a steep gradient ($\sim -0.09\,\rm{dex\,kpc^{-1}}$) and high $\sigma_{\rm{g}}$ ($\sim 109\,\rm{km\,s^{-1}}$). \autoref{fig:grad_vsigma} shows the same data as in \autoref{fig:grad_sigma}, but as a function of $v_{\phi}/\sigma_{\rm{g}}$, thereby separating galaxies that are rotation-dominated from those that are dispersion-dominated. The main conclusion that we can draw from \autoref{fig:grad_vsigma} is that all dispersion-dominated galaxies ($v_{\phi}/\sigma_{\rm{g}} \lesssim 1$) possess shallow or flat gradients, whereas rotation-dominated galaxies ($v_{\phi}/\sigma_{\rm{g}} \gtrsim 1$) show a large scatter, and can have flat as well as steep gradients.\footnote{Note, however, that this conclusion depends on the value $v_{\phi}/\sigma_{\rm{g}}\sim 1$ that we choose to separate rotation- and dispersion-dominated galaxies. If we were to use $v_{\phi}/\sigma_{\rm{g}} = 3$ as the break-point, for example, we would find that dispersion-dominated galaxies show a large scatter in metallicity gradients whereas rotation-dominated galaxies show shallow or flat gradients. A more precise statement, which we will see below is naturally predicted by our theoretical model, is that metallicity gradients show a very large scatter for $v_\phi/\sigma_{\rm g}\sim\mbox{few}$, and shallow gradients on either side of this region.} The exception to this is a couple of dispersion-dominated galaxies at $z < 1$ in the \textit{MUSE-WIDE} sample that exhibit steep gradients. We also find from \autoref{fig:grad_vsigma} that the scatter in gradients narrows down as $v_{\phi}/\sigma_{\rm{g}}$ increases. While our data compilation spans a wide redshift range, the above conclusions are not substantially different when considering just $z>1$ versus $z<1$. A more complete analysis exploring possible evolutionary effects requires a larger dataset.

Recent cosmological simulations like FIRE \citep{{2014MNRAS.445..581H,2018MNRAS.480..800H}} and IllustrisTNG50 \citep{2018MNRAS.473.4077P} have also explored the connection between metallicity gradients and kinematics, particularly focusing on the relation between gradients and $v_{\phi}/\sigma_{\rm{g}}$. One of the key results of both these simulations is that negative metallicity gradients only form in galaxies with $v_{\phi}/\sigma_{\rm{g}} > 1$ (\textit{i.e.,} rotation-dominated systems), however, many such galaxies also show shallow gradients \citep{2017MNRAS.466.4780M,2020arXiv200710993H}. These simulations also find that dispersion-dominated galaxies always show shallow/flat gradients, consistent with mixing due to efficient feedback. However, some simulations may require more powerful radial mixing or feedback to match both kinematics and gradients \citep{2013A&A...554A..47G}. We see from \autoref{fig:grad_sigma} and \autoref{fig:grad_vsigma} that these findings are consistent with the data analysed here, with only a couple of dispersion-dominated outliers that show steep metallicity gradients.

\section{Comparison with analytic model for metallicity gradients}
\label{s:compare_theory}
In order to better understand the underlying physics that drives the diversity of metallicity gradients found in high-redshift galaxies, we compare the observations with the analytic metallicity gradient model we presented in \cite{2020aMNRAS.xxx..xxxS}. Our model predicts the radial distribution of gas phase metallicities based on the equilibrium between production, consumption, loss and transport of metals in galaxies. It is a standalone metallicity model, but requires inputs from a galaxy evolution model to describe the properties of the gas -- velocity dispersion, surface densities of gas and star formation, etc. -- to solve for the metallicity. We use the galactic disc model of \cite{2018MNRAS.477.2716K} for this purpose, since we showed in previous works that using this model allows us to successfully reproduce the observed trend of metallicity gradient with redshift \citep{2020aMNRAS.xxx..xxxS}, as well as the mass--metallicity and mass--metallicity gradient relations (MZR and MZGR) found in local galaxies \citep{2020bMNRAS.xxx..xxxS}. Note that, in what follows, we do not \textit{fit} the model to the data while comparing the two.

\subsection{Model description}
\label{s:compare_theory_model}
In our model, the metallicity distribution profile in the galactic disc depends on four dimensionless ratios \citep[equations 13, 37 -- 40 in][]{2020aMNRAS.xxx..xxxS},
\begin{eqnarray}
\mathcal{T} \propto \left(\frac{v_{\phi}}{\sigma_{\rm{g}}}\right)^2\,\,\mathrm{\left[metal\,\, equilibrium\right]}, \\
\mathcal{P} \propto \left(1 - \frac{\sigma_{\mathrm{sf}}}{\sigma_{\rm{g}}}\right)\,\,\mathrm{\left[metal\,\, advection\right]}, \\
\mathcal{S} \propto \phi_y\left(\frac{v_{\phi}}{\sigma_{\rm{g}}}\right)^2 \,\,\mathrm{\left[metal\,\, production\right]},\\
\mathcal{A} \propto \frac{1}{\sigma^3_g} \,\,\mathrm{\left[cosmic\,\, accretion\right]}\,,
\end{eqnarray}
where we have only retained the dependencies on $v_{\phi}$ and $\sigma_{\rm{g}}$ for the purposes of the present study. Here, $\mathcal{T}$ is the ratio of the orbital to diffusion timescales, which describes the time it takes for a given metallicity distribution to reach equilibrium, $\mathcal{P}$ is the Péclet number of the galaxy \citep[e.g.,][]{patankar1980numerical}, which describes the ratio of advection to diffusion of metals in the disc, $\mathcal{S}$ is the source term, which describes the ratio of metal production (including loss of metals in outflows) and diffusion, and $\mathcal{A}$ is the ratio of cosmic accretion (infall) to diffusion. Finally, $\sigma_{\rm{sf}}$ denotes the velocity dispersion that can be maintained by star formation feedback alone, with no additional energy input from transport of gas through the disc \citep{2018MNRAS.477.2716K}. Note that the model can only be applied in cases where the metal equilibration time (dictated by $\mathcal{T}$) is shorter than the Hubble time and shorter than or comparable to the molecular gas depletion time. If these conditions are not met, the metallicity distribution does not reach equilibrium within the galaxy. We showed in \citet[Section 5]{2020aMNRAS.xxx..xxxS} that inverted gradients may or may not be in equilibrium, so for this work we do not apply our model to study such gradients. 

The parameter $\phi_y$ that appears in $\mathcal{S}$ describes the reduced yield of metals in the disc due to preferential metal ejection through galactic outflows: $\phi_y=1$ corresponds to metals being thoroughly-mixed into the interstellar medium (ISM) before ejection, whereas $\phi_y=0$ means that all the newly-produced metals are directly ejected before they can mix into the ISM. In line with previous works, we leave $\phi_y$ as a free parameter in the model. However, we showed in \cite{2020bMNRAS.xxx..xxxS} that our model reproduces both the local MZR and the MZGR only if $\phi_y$ increases with $M_{\star}$: low-mass galaxies prefer a lower $\phi_y$, and vice-versa.

\begin{figure*}
\includegraphics[width=1.0\columnwidth]{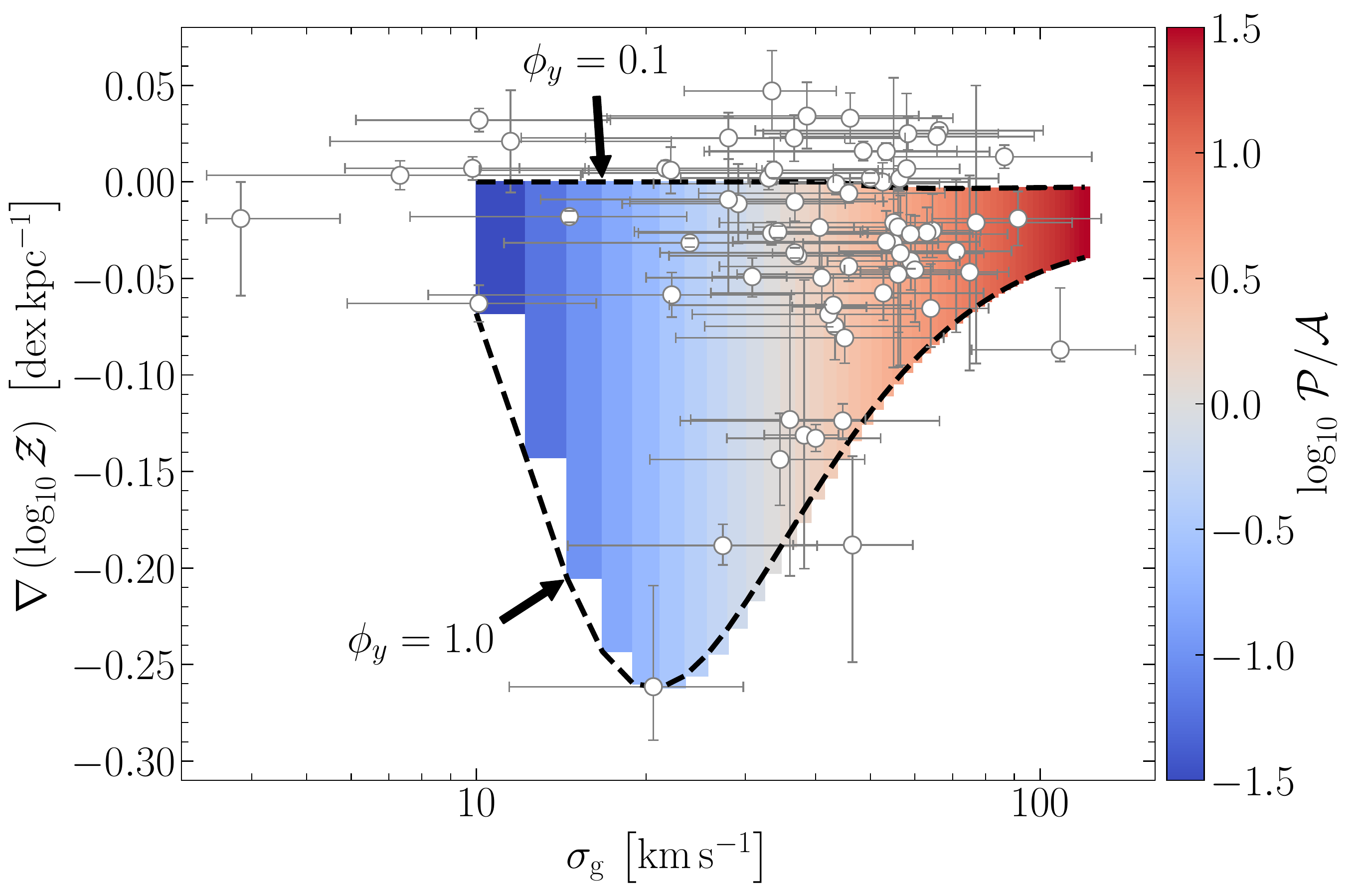}
\includegraphics[width=1.0\columnwidth]{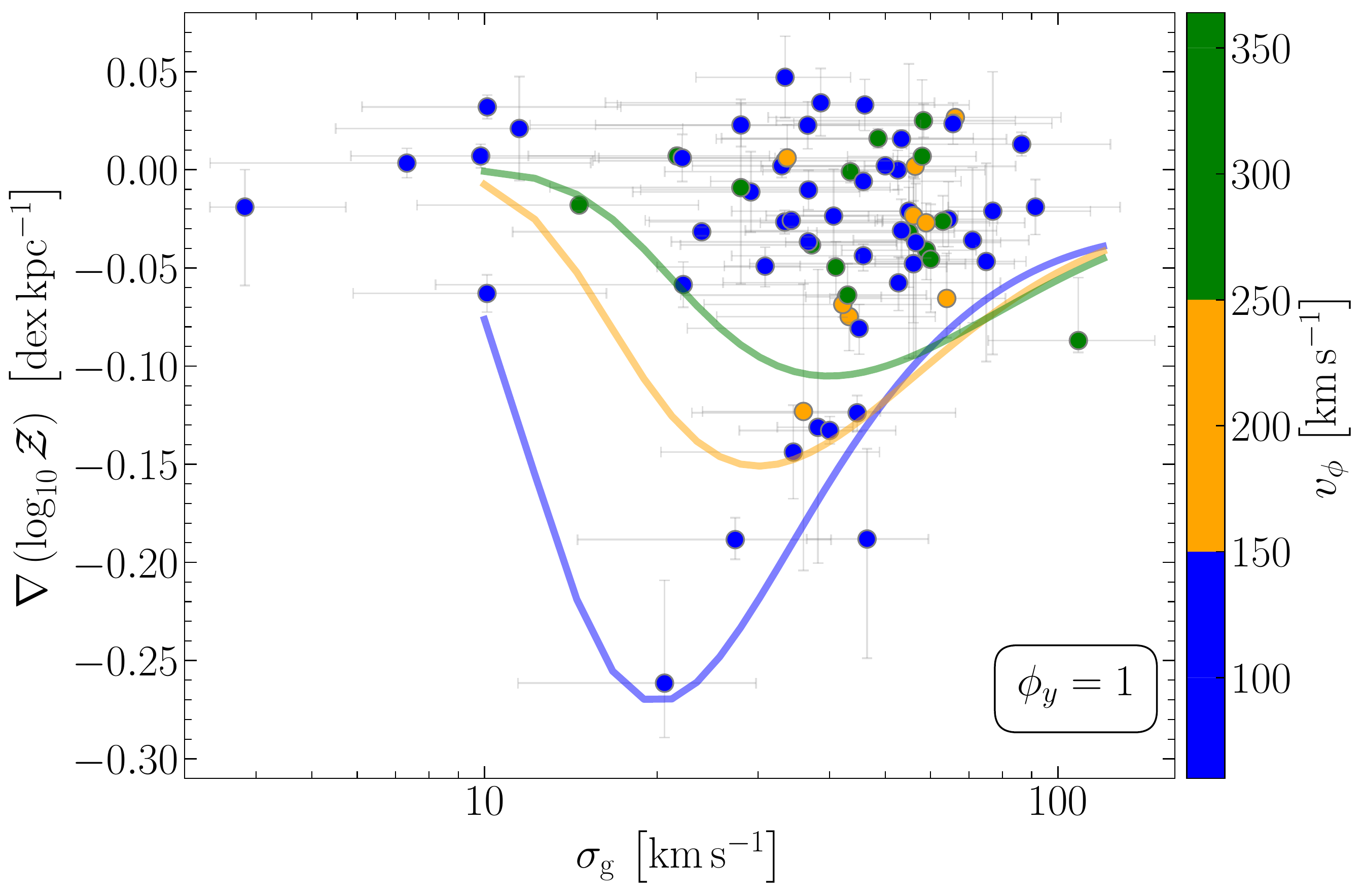}
\caption{\textit{Left panel:} Same data as \autoref{fig:grad_sigma}, but overplotted with one of the \protect\cite{2020aMNRAS.xxx..xxxS} models. The model is for a high-redshift galaxy at fixed $v_{\phi}=105\,\rm{km\,s^{-1}}$ (median $v_{\phi}$ in the data) and $z=2$. The spread in the model (represented by the length of the colored bands) is a result of the yield reduction factor $\phi_y$, which describes the preferential ejection of metals through galactic winds. Here we show models with $\phi_y=0.1$--$1$, where the top and bottom dashed lines corresponds to $\phi_y=0.1$ and $1.0$, respectively. The colorbar denotes the ratio of advection of gas ($\mathcal{P}$) to cosmic accretion of metal-poor gas ($\mathcal{A}$). The steepest gradients produced by the model correspond to a transition from the accretion-dominated to the advection-dominated regime, as $\sigma_{\rm{g}}$ increases. \textit{Right panel:} Same as the left panel, but overlaid with different models (corresponding to different $v_{\phi}$) at $z=2$. Only the $\phi_y=1$ model is shown here; thus, the model curves represent the most negative gradients produced by the model for a given set of parameters. The data are also binned around the model $v_{\phi}$ as shown through the colorbar. Note that the model is not being fit to the data.}
\label{fig:grad_sigma_model}
\end{figure*}

\begin{figure}
\includegraphics[width=1.0\columnwidth]{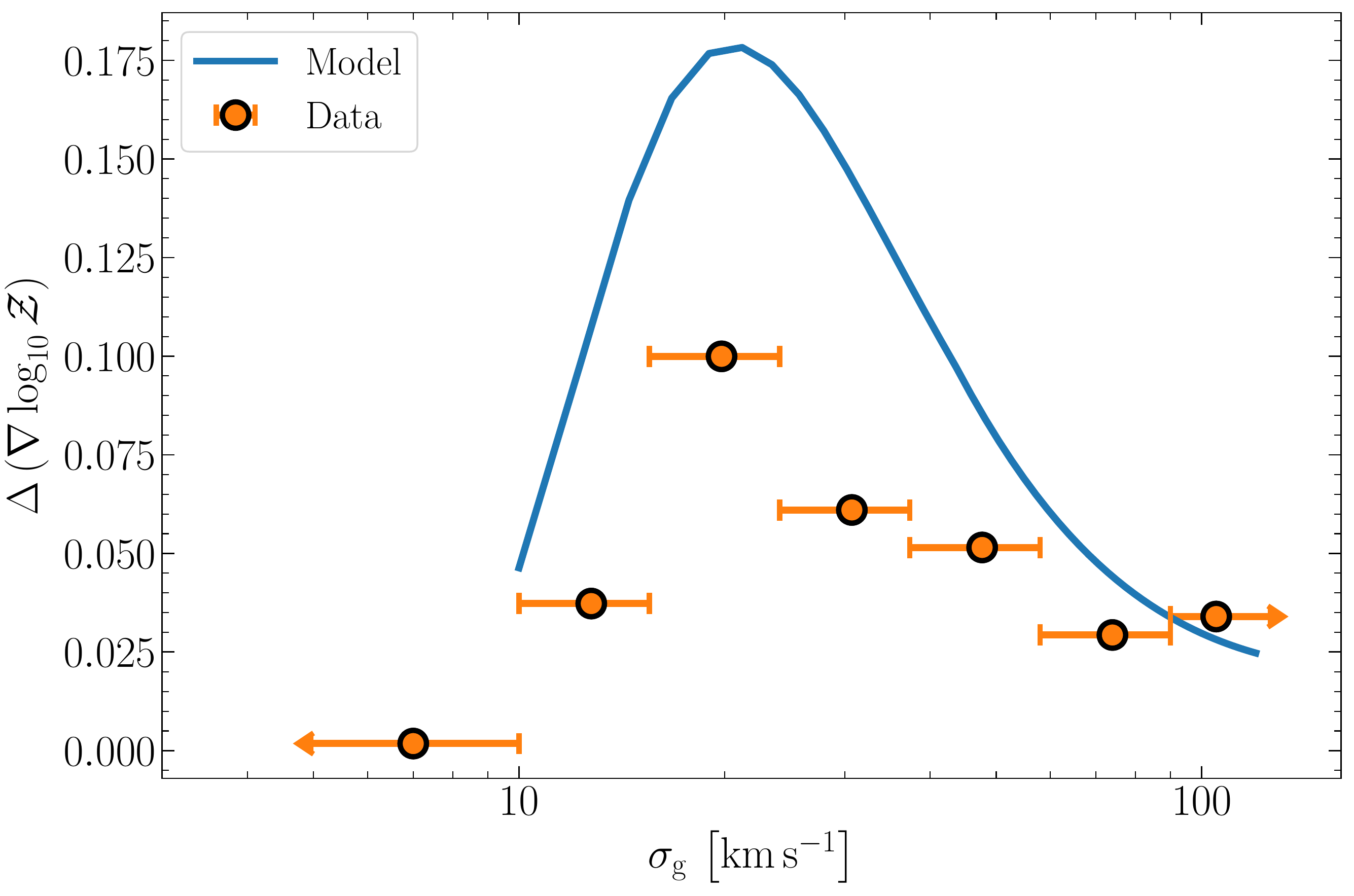}
\caption{Scatter in the model metallicity gradient shown in the left panel of \autoref{fig:grad_sigma_model}, compared to the scatter for observed data, as a function of the gas velocity dispersion $\sigma_{\rm{g}}$ for a fixed rotational velocity $v_{\phi}=105\,\rm{km\,s^{-1}}$ (median $v_{\phi}$ in the data). The scatter in the model is calculated as 68 per cent of the difference between the model metallicity gradients with $\phi_y = 1$ and $0.1$ at every $\sigma_{\rm{g}}$. Errors on the scatter in the data represent the width of the bins used.}
\label{fig:grad_sigma_scatter}
\end{figure}

\subsection{Model application}
\label{s:compare_theory_application}
To produce metallicity gradients from the model, we select a value of $v_{\phi}$ and $\sigma_{\rm{g}}$, and fix all other parameters in the model to those appropriate for high-$z$ galaxies \citep[Tables~1 and~2]{2020aMNRAS.xxx..xxxS} at $z=2$; we discuss how the results depend on the choice of redshift below. We find the spatial distribution of metallicity, $\mathcal{Z}(r)$, within $0.5-2.0\,r_{\rm{e}}$ using equation~41 of \citet{2020aMNRAS.xxx..xxxS}, which we then linearly fit in logarithmic space to obtain a metallicity gradient in $\rm{dex\,kpc^{-1}}$ from the model \citep[e.g.,][]{2018MNRAS.478.4293C}. We use this range in $r$ because it is well matched to the observations ($r_{\mathrm{vel}}/r_{\mathrm{e}}\approx 2$) and the input galaxy model does not apply to the innermost regions of the galaxy. 

While the choice of most of the parameters used as inputs into the metallicity model have no appreciable effect on the results, some (e.g., the Toomre $Q$ parameter, and the circumgalactic medium metallicity $\mathcal{Z}_{\rm{CGM}}$) matter at the level of tens of percent. For example, changing the Toomre $Q$ parameter by a factor of 2 induces a 20 percent change in the metallicity gradient. The effects of changing the Toomre $Q$ are similar to that of changing the rotation curve index $\beta$ in the model, a topic we explore below in \autoref{s:grad_beta}. Changing $\mathcal{Z}_{\rm{CGM}}$ by $\pm0.1$ changes the metallicity gradient by at most $\pm$50 percent. This implies that increasing the CGM metallicity leads to shallower metallicity gradients. Finally, varying the redshift at which we compute the gradient by $\pm1$ yields changes in the gradient from $\mp36$ per cent for massive galaxies to $\mp19$ percent for low-mass galaxies. However, the overall impact of these parameters on the resulting metallicity gradients is limited compared to the dependence on $\phi_y$, so in the following we focus on studying the effects of changing $\phi_y$. In the main text that follows, we will continue to measure metallicity gradients in $\rm{dex\,kpc^{-1}}$; we provide results on metallicity gradients measured in $\mathrm{dex}\,r_{\rm{e}}^{-1}$, which would potentially account for evolution in galaxy size, in \aref{s:app_galaxysize}.

\subsubsection{Metallicity gradient versus velocity dispersion}
\label{s:grad_sigma}
The left-hand panel of \autoref{fig:grad_sigma_model} shows the same observational data as in \autoref{fig:grad_sigma}, now with our model as computed for a fixed $v_{\phi}=105\,\mathrm{km\,s^{-1}}$ (the median $v_{\phi}$ in the data). Since $\phi_y$ is a free parameter, we obtain a range of model predictions at every $\sigma_{\rm{g}}$; the range shown in the plot corresponds to varying $\phi_y$ between 0.1 and 1, as represented by the arrows on \autoref{fig:grad_sigma_model}. We color-code bars within this range by the ratio $\mathcal{P}/\mathcal{A}$, which describes the relative importance of advection and accretion of metal-poor gas. A key conclusion that can be drawn from this model--data comparison is that the model predicts flat metallicity gradients in galaxies with high $\sigma_{\rm{g}}$, irrespective of $\phi_y$, in good agreement with the observational data. The model generates a uniform metallicity distribution across the disc (\textit{i.e.,} a flat/shallow gradient) as a result of efficient radial transport of the gas. The model does not produce any steep negative ($< -0.1\,\rm{dex\,kpc^{-1}}$) metallicity gradients at high $\sigma_{\rm{g}}$ ($>60\,\rm{km\,s^{-1}}$), consistent with both data and simulations.

The largest diversity in metallicity gradients in the model occurs at $\sigma_{\rm{g}} \approx 20-40\,\rm{km\,s^{-1}}$, where galaxies transition from being accretion-dominated (blue, $\mathcal{P}<\mathcal{A}$) to being advection-dominated (red, $\mathcal{P}>\mathcal{A}$). This transition in the ratio $\mathcal{P}/\mathcal{A}$ and the corresponding scatter in the steepness of the metallicity gradients are key results of the model. Moreover, the transition in $\mathcal{P}/\mathcal{A}$ from high to low values mirrors the transition seen in $\sigma_{\rm{g}}$ from gravity-driven to star formation feedback-driven turbulence \citep{2018MNRAS.477.2716K}. The region around the transition is where both advection ($\mathcal{P}$) and accretion ($\mathcal{A}$) are weaker as compared to metal production ($\mathcal{S}$), resulting in steep metallicity gradients, since star formation and thus metal production are centrally peaked \citep{2018MNRAS.477.2716K}\footnote{Note that the input cosmic accretion profile in the model is also centrally peaked, similar to the SFR profile. However, as we show in \citep[Appendix~A]{2020aMNRAS.xxx..xxxS}, changing the form of the input accretion profile has only modest effects on the resulting metallicity gradients: less centrally-peaked accretion profiles give rise to slightly steeper gradients.}. The scatter near the transition arises due to the yield reduction factor $\phi_y$, which can decrease the strength of $\mathcal{S}$ as compared to $\mathcal{P}$ or $\mathcal{A}$ because $\mathcal{S} \propto \phi_y$. Lastly, we note that while metal diffusion is an important process that can also flatten the gradient, it never simultaneously dominates advection and cosmic accretion, since both $\mathcal{P}$ and $\mathcal{A}$ are never less than unity at the same time.

The model is consistent with the very few data points at low $\sigma_{\rm{g}}$, which show shallow/flat metallicity gradients. In the model, the flattening at the low-$\sigma_{\rm{g}}$ end is caused by accretion of metal-poor gas, following a $1/r^2$ profile, that dilutes the metallicity primarily in the central regions. Given the scarcity of data at low $\sigma_{\rm{g}}$, as well as significant observational uncertainties, it is unclear whether the trend seen in the model is also present in the data. Future instruments with higher sensitivity and spectral and spatial resolution (e.g., GMTIFS, HARMONI, MAVIS, ERIS) will be able to measure low $\sigma_{\rm{g}}$ in high-redshift galaxies with higher precision \citep{2014SPIE.9147E..25T,2017PASA...34...53F,2018SPIE10702E..09D,2020MNRAS.498.1891R,2020arXiv200909242M,2020SPIE11447E..A0E}, expanding the currently available sample by a considerable margin.

In the right-hand panel of \autoref{fig:grad_sigma_model}, we now fix $\phi_y=1$ (implying no preferential metal ejection in winds), and look at the model differences for different values of $v_{\phi}$. Note that $v_{\phi}$ is a proxy for stellar mass, as higher $v_{\phi}$ typically corresponds to massive galaxies in the compiled sample. The data are the same as in the left-hand panel of \autoref{fig:grad_sigma_model}, but now binned and color-coded by the measured $v_{\phi}$. Thus, the model curves represent the steepest metallicity gradients that we can obtain for the given set of galaxy parameters. We emphasize that we do not fit the model to the data while plotting the model curves. It is clear that low $v_{\phi}$ (low mass) galaxies show more scatter in the model gradients as compared to high $v_{\phi}$ (massive) galaxies, consistent with observations \citep{2018MNRAS.478.4293C,2020arXiv201103553S}. As $v_{\phi}$ increases, the point of inflection (or, the point of steepest gradients) shifts toward higher $\sigma_{\rm{g}}$ and toward shallower metallicity gradients. Additionally, for sufficiently high $\sigma_{\rm{g}}$, models with different $v_{\phi}$ converge towards a lower bound for metallicity gradients, implying that the flatness of metallicity gradients at high $\sigma_{\rm{g}}$ is independent of the galaxy mass (see, however, \autoref{s:grad_beta}). When the data are binned in $v_{\phi}$, they are broadly consistent with the model. Thus, the model suggests a lower limit in metallicity gradients at high velocity dispersions consistent with the compiled data.

Taken at face value, it seems from the left panel of \autoref{fig:grad_sigma_model} that most galaxies in the sample favor a value of $\phi_y$ close to 0.1, implying a high metal enrichment in their winds. A close examination of the right panel of \autoref{fig:grad_sigma_model} reveals that this is only the case for galaxies with $v_{\phi} < 150\,\rm{km\,s^{-1}}$ (\textit{i.e.,} low-mass galaxies); galaxies with higher $v_{\phi}$ prefer both low and high values of $\phi_y$. When combined with results from the local Universe showing that low-mass galaxies prefer lower $\phi_y$ \citep{2020bMNRAS.xxx..xxxS}, the finding that low-mass galaxies at high redshift also prefer lower $\phi_y$ is not surprising. Not only were outflows more common in the past in actively star-forming galaxies \citep{2015MNRAS.454.2691M}, galaxies also had shallower potential wells \citep{2010ApJ...710..903M} that made it easier for metals to escape via galactic winds without mixing into the ISM. This effect is more pronounced at the low-mass end, thus low-mass galaxies tend to prefer $\phi_y \sim 0.1$.

\autoref{fig:grad_sigma_scatter} quantifies the scatter present in the model and the data as a function of $\sigma_{\rm{g}}$. To construct a $1\sigma$ scatter in the model in the absence of \textit{a priori} knowledge of the distribution of $\phi_y$, we simply compute  model gradients for $\phi_y=1$ and $0.1$ for every $\sigma_{\rm{g}}$, and take the scatter to be 68 per cent of this range, \textit{i.e.,} our model-predicted ``scatter'' is simply 68 per cent of the distance between the two black dashed lines in the left panel of \autoref{fig:grad_sigma_model}. To estimate the scatter in the data, we bin by $\sigma_{\rm{g}}$ such that we have one bin each for $\sigma_{\rm{g}} < 10\,\rm{km\,s^{-1}}$ and $\sigma_{\rm{g}} > 90\,\rm{km\,s^{-1}}$, and we divide the parameter space $10 \leq \sigma_{\rm{g}}/\rm{km\,s^{-1}} \leq 90$ in five logarithmically spaced bins.

\begin{figure*}
\includegraphics[width=1.0\columnwidth]{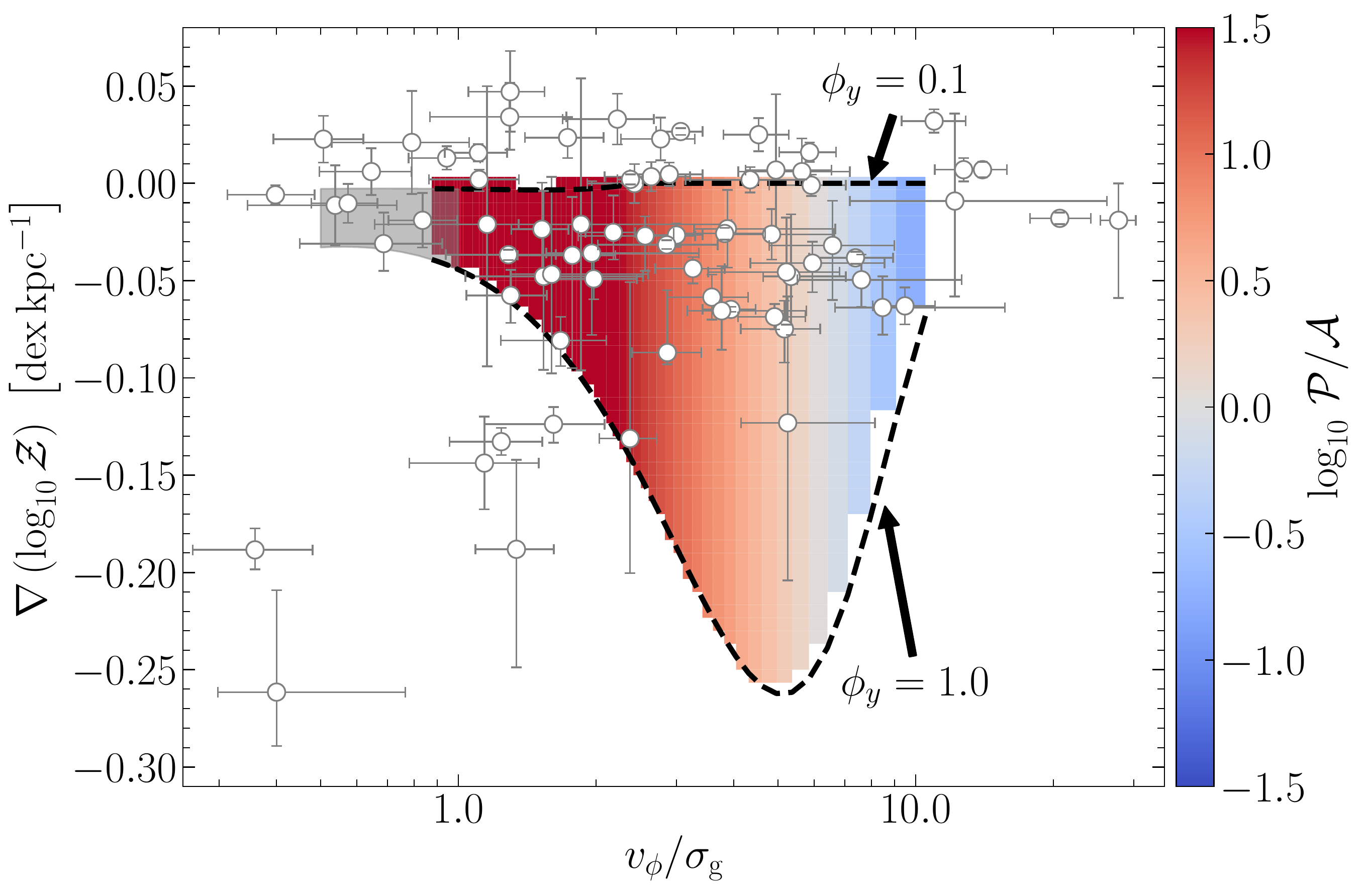}
\includegraphics[width=1.0\columnwidth]{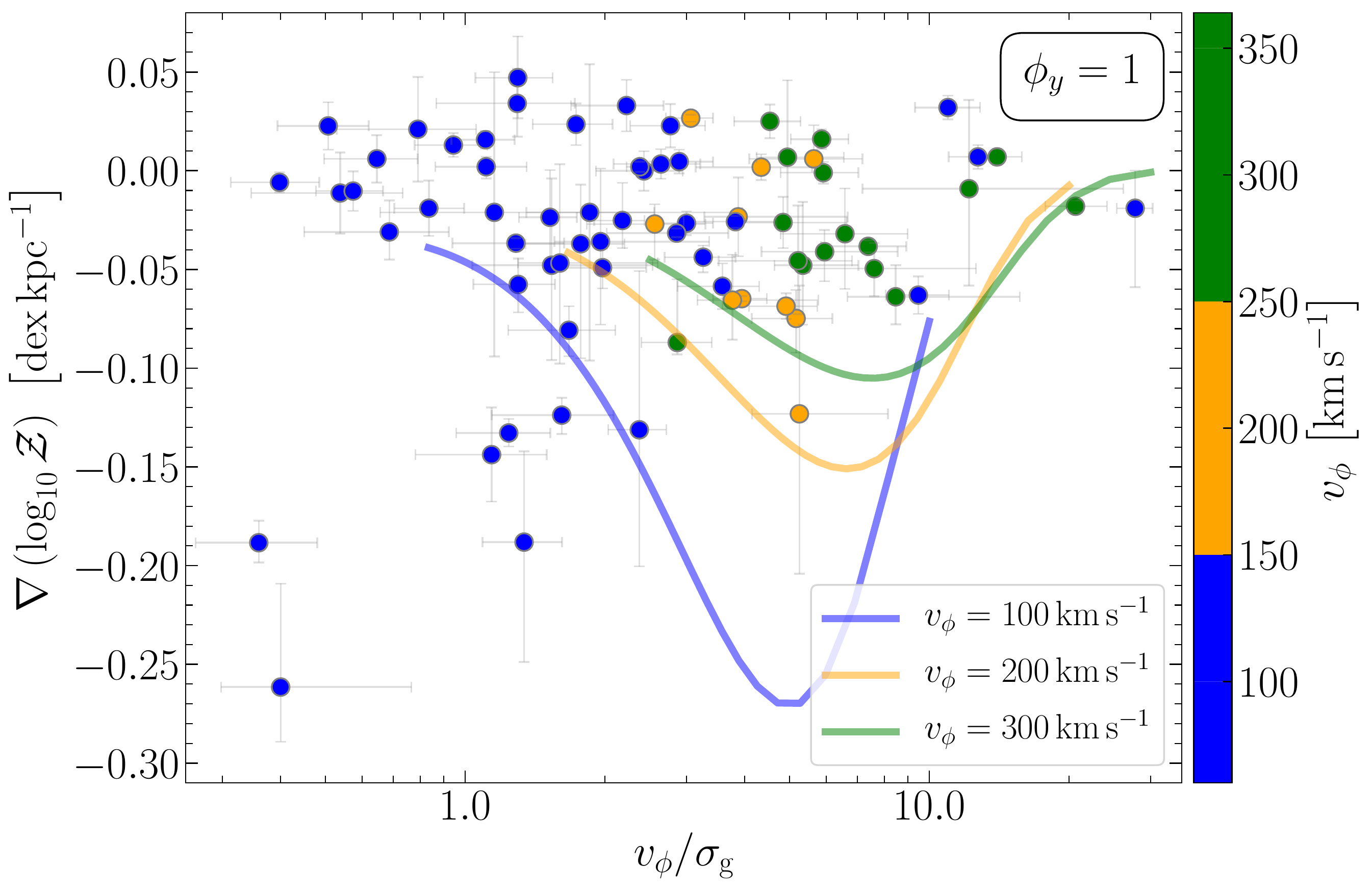}
\caption{\textit{Left panel:} Same data as in \autoref{fig:grad_vsigma}, and model as in \autoref{fig:grad_sigma_model} (left panel), but now plotted as a function of $v_{\phi}/\sigma_{\rm{g}}$ for a fixed $v_{\phi}=105\,\rm{km\,s^{-1}}$ (median $v_{\phi}$ in the data). The grey-shaded area corresponds to the predictions of the model for $v_{\phi}/\sigma_{\rm{g}} < 1$, where the assumption of a disc-like structure likely breaks down, hence the galaxy disc model \protect\citep{2018MNRAS.477.2716K} used as an input to the metallicity model \protect\citep{2020aMNRAS.xxx..xxxS} may not be fully applicable. \textit{Right panel:} Same as \autoref{fig:grad_sigma_model} (right panel), but with metallicity gradients plotted as a function of $v_{\phi}/\sigma_{\rm{g}}$, overlaid with a set of models for different $v_{\phi}$. The models are not fit to the data.}
\label{fig:grad_vsigma_model}
\end{figure*}

\begin{figure}
\includegraphics[width=1.0\columnwidth]{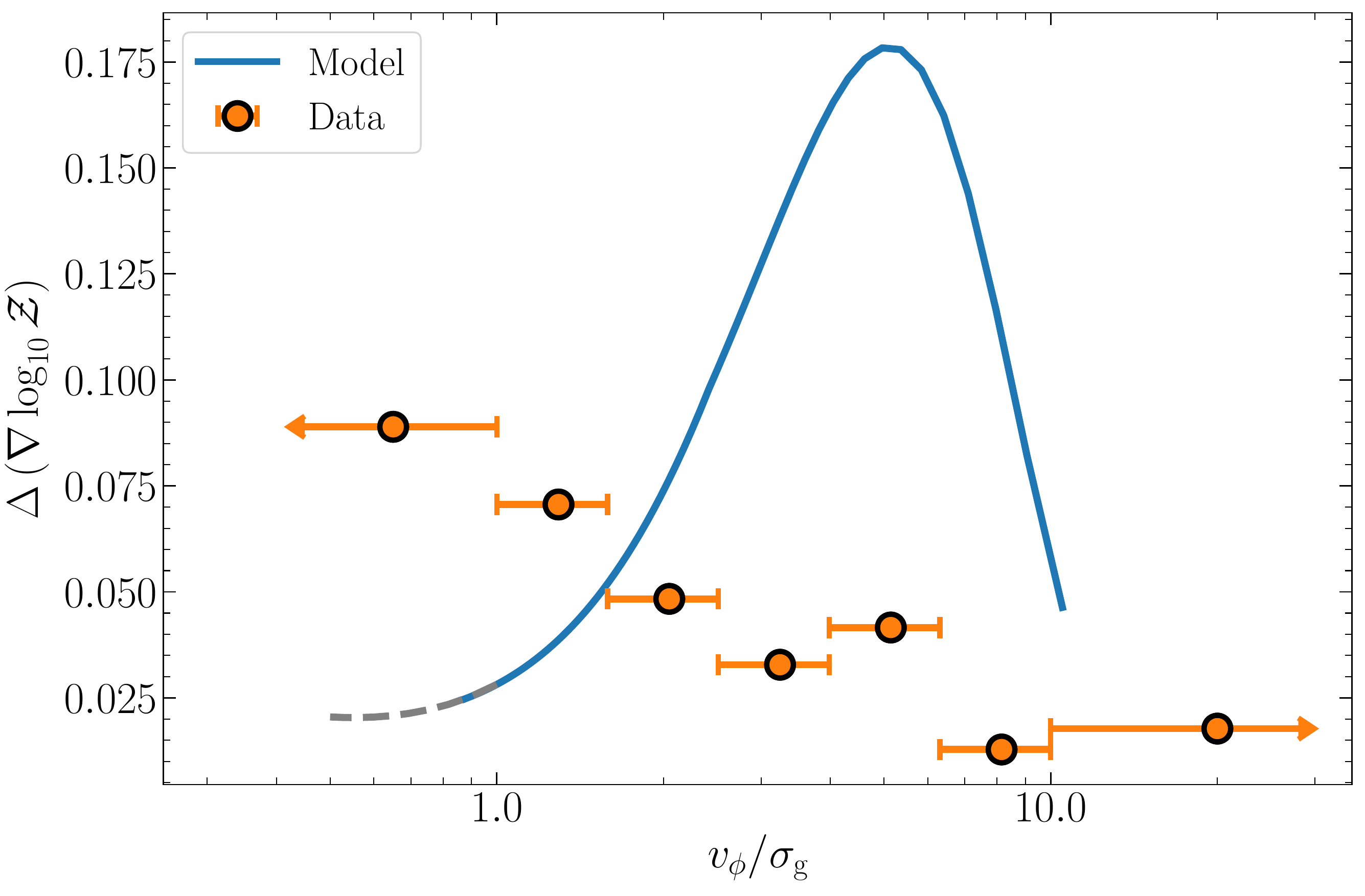}
\caption{Scatter in the model metallicity gradient shown in the left panel of \autoref{fig:grad_vsigma_model}, compared to the scatter for observed data, as a function of rotational support, $v_{\phi}/\sigma_{\rm{g}}$. The scatter in the model is calculated as 68 per cent of the difference between the model metallicity gradients with $\phi_y = 1$ and $0.1$ at every $\sigma_{\rm{g}}$ for a fixed $v_{\phi}=105\,\rm{km\,s^{-1}}$ (median $v_{\phi}$ in the data). The grey-shaded extension of the model scatter corresponds to the grey-shaded region in the left panel of \autoref{fig:grad_vsigma_model}. Errors on the scatter in the data represent the width of the bins used.}
\label{fig:grad_vsigma_scatter}
\end{figure}

We find from \autoref{fig:grad_sigma_scatter} that the model is able to reproduce the qualitative shape of the variation in scatter with $\sigma_{\rm g}$ observed in the data, but not the exact level of scatter. This is not surprising, since the true scatter expected for the model depends on the distribution of $\phi_y$ values in real galaxies, which is at present unknown. Future observations of galactic wind metallicity at low and high redshifts, or simulations with enough resolution to capture the hot-cold interface where mixing between SN ejecta and the ambient ISM occurs (something no current cosmological simulation possesses -- \citealt{Gentry19a}), will enable us to constrain $\phi_y$ and provide a more quantitative analysis of how $\phi_y$ scales with galaxy mass at different redshifts.

\subsubsection{Metallicity gradient versus rotational support}
\label{s:grad_vsigma}
The ratio of rotation to velocity dispersion provides a quantification of the overall rotational support of a galaxy. The left-hand panel of \autoref{fig:grad_vsigma_model} shows the metallicity gradients as a function of $v_{\phi}/\sigma_{\rm{g}}$ in the data, overplotted with the analytic model for fixed $v_{\phi}=105\,\rm{km\,s^{-1}}$. The parameter space of the model denotes variations in $\mathcal{P}/\mathcal{A}$, the same as that shown in \autoref{fig:grad_sigma_model}. The grey-shaded region corresponds to an extrapolation of the model where it is not directly applicable because the assumption of a disc likely breaks down at $v_{\phi}/\sigma_{\rm{g}}<1$. We first find a steepening of the gradient in the model as $v_{\phi}/\sigma_{\rm{g}}$ increases from $< 1$ to $\sim 10$, after which the gradients begin to flatten again for  $v_{\phi}/\sigma_{\rm{g}} \gtrsim 10$. We can again understand this trend in terms of $\mathcal{P}/\mathcal{A}$: values of $v_{\phi}/\sigma_{\rm{g}} \gtrsim 10$ typically correspond to massive galaxies, within which strong centrally peaked accretion (large $\mathcal{A}$) flattens the gradients. Galaxies with $v_{\phi}/\sigma_{\rm{g}} \lesssim 1$ have flat gradients due to strong advection of gas through the disc (large $\mathcal{P}$) mixing and therefore homogenising the metal distribution throughout the disc. In the intermediate range of $v_{\phi}/\sigma_{\rm{g}}$, the gradients are the steepest because the production term $\mathcal{S}$ dominates over both $\mathcal{P}$ and $\mathcal{A}$.

The location of the turnover is sensitive to the value of $v_{\phi}$, as we show in the right-hand panel of \autoref{fig:grad_vsigma_model}. This figure is similar to the right-hand panel of \autoref{fig:grad_sigma_model}, but we now show metallicity gradients as a function of $v_{\phi}/\sigma_{\rm{g}}$ for different values of $v_{\phi}$. We see that as $v_{\phi}$ increases, the parameter space of the model shifts to flatter gradients and higher $v_{\phi}/\sigma_{\rm{g}}$. These shifts in the inflection point where galaxies transition from the advection-dominated to the accretion-dominated regime imply that massive galaxies have higher $v_{\phi}/\sigma_{\rm{g}}$ and shallower gradients as compared to low-mass galaxies. 

Similar to \autoref{fig:grad_sigma_model}, we notice that most low-mass (low-$v_{\phi}$) galaxies prefer a lower value of $\phi_y$. The bounds provided by the model in terms of the most negative gradient it can produce (represented by $\phi_y=1$) are consistent with the majority of the data. The four rotation-dominated ($v_{\phi}/\sigma_{\rm{g}}>1$) outliers that we observe have $v_{\phi}$ less than $100\,\rm{km\,s^{-1}}$, so it is not surprising that the model does not produce a bound that is consistent with these galaxies. However, as we saw earlier in the right panel of \autoref{fig:grad_sigma_model}, the same outliers are within the constraints of the model when we study the trends with $\sigma_{\rm{g}}$. Thus, while the agreement of the model with the lower bound in metallicity gradient as a function of $\sigma_{\rm{g}}$ is good, it is less so in metallicity gradient as a function of $v_{\phi}/\sigma_{\rm{g}}$. This is a shortcoming of the model, which may be a result of our fundamental approach of treating galaxies as discs breaking down as we approach $v_{\phi}/\sigma_{\rm{g}} \sim 1$, or restricting $v_{\phi}$ to a handful of values in the model.

\autoref{fig:grad_vsigma_scatter} plots the scatter in the model and the data as a function of $v_{\phi}/\sigma_{\rm{g}}$, in the same manner as that in \autoref{fig:grad_sigma_scatter}. We bin the data such that we have one bin each for $v_{\phi}/\sigma_{\rm{g}} < 1$ and $v_{\phi}/\sigma_{\rm{g}} > 10$, with five logarithmically spaced bins in between. Consistent with our findings above, we see that the model fails to reproduce the shape of the scatter as a function of rotational support. The discrepancy is largely due to restricting the model to a single value of $v_{\phi}$ whereas the data spans a wide range in $v_{\phi}$. However, the current sample is too limited for us to bin the data in different $v_{\phi}$ bins and study the trends in the scatter by using several different values of $v_{\phi}$ in the model.

Overall, we find that the model is able to reproduce the observed non-monotonic trends (but not the scatter) between metallicity gradients and $v_{\phi}/\sigma_{\rm{g}}$, and provide a physical explanation for them. However, reproducing the full distribution of the data is beyond the scope of the model without better constraints on model parameters like $\phi_y$. Additional data, particularly at high mass ($M_{\star} \sim 10^{10.5}\,\rm{M_{\odot}}$) and low redshift ($0 < z < 1$) would provide further constraints on the performance of the model as a function of rotational support \citep[e.g.,][]{2020arXiv201113567F}.

\subsubsection{Metallicity gradient versus rotation curve index}
\label{s:grad_beta}
So far, we have only considered applications of the model that assume a flat rotation curve, $\beta=0$, for all galaxies, where $\beta \equiv d\ln v_\phi/d\ln r$ is the index of the rotation curve. However, at high redshift when galaxies are more compact, the visible baryons are more likely to be a in a baryon-dominated regime, which can give rise to non-flat rotation curves such that $\beta \neq 0$. Recent observations suggest that the inner regions of several high-$z$ galaxies are baryon dominated (\citealt{2017Natur.543..397G,2020ApJ...902...98G,2017ApJ...840...92L,2018ApJ...854L..28T}; see, however, \citealt{2019MNRAS.485..934T}), such that $\beta < 0$. Keeping these findings in mind, we now explore the effects of varying $\beta$ on the metallicity gradients produced by our model.

In the context of our model, the rotation curve has several effects. First, the model is based on the premise that Toomre $Q\approx 1$, and $Q$ depends on the epicyclic frequency and thus on $\beta$ -- changing $\beta$ therefore changes the relationship between the gas surface density and the velocity dispersion; this manifests as a change in the source term $\mathcal{S}$, which depends on the star formation rate and thus on the gas content. Second, the rotation curve index changes the amount of energy released by inward radial flows, which alters the inflow rate required to maintain energy balance; this manifests as a change in $\mathcal{P}$.\footnote{Both of these effects also alter the equilibration timescale, and thus $\mathcal{T}$, but by little enough that our finding that all the galaxies under consideration are in equilibrium is unaffected. We therefore do not discuss $\mathcal{T}$ further.} From \citet[equations~38 and~39]{2020aMNRAS.xxx..xxxS}, we find that $\mathcal{P} \propto (1+\beta)/(1-\beta)$ and $\mathcal{S} \propto (1+\beta)$. Thus, $\beta < 0$ reduces both $\mathcal{S}$ and $\mathcal{P}$, weakening metal production and advection in comparison to cosmological accretion and diffusion.

\begin{figure}
\includegraphics[width=\columnwidth]{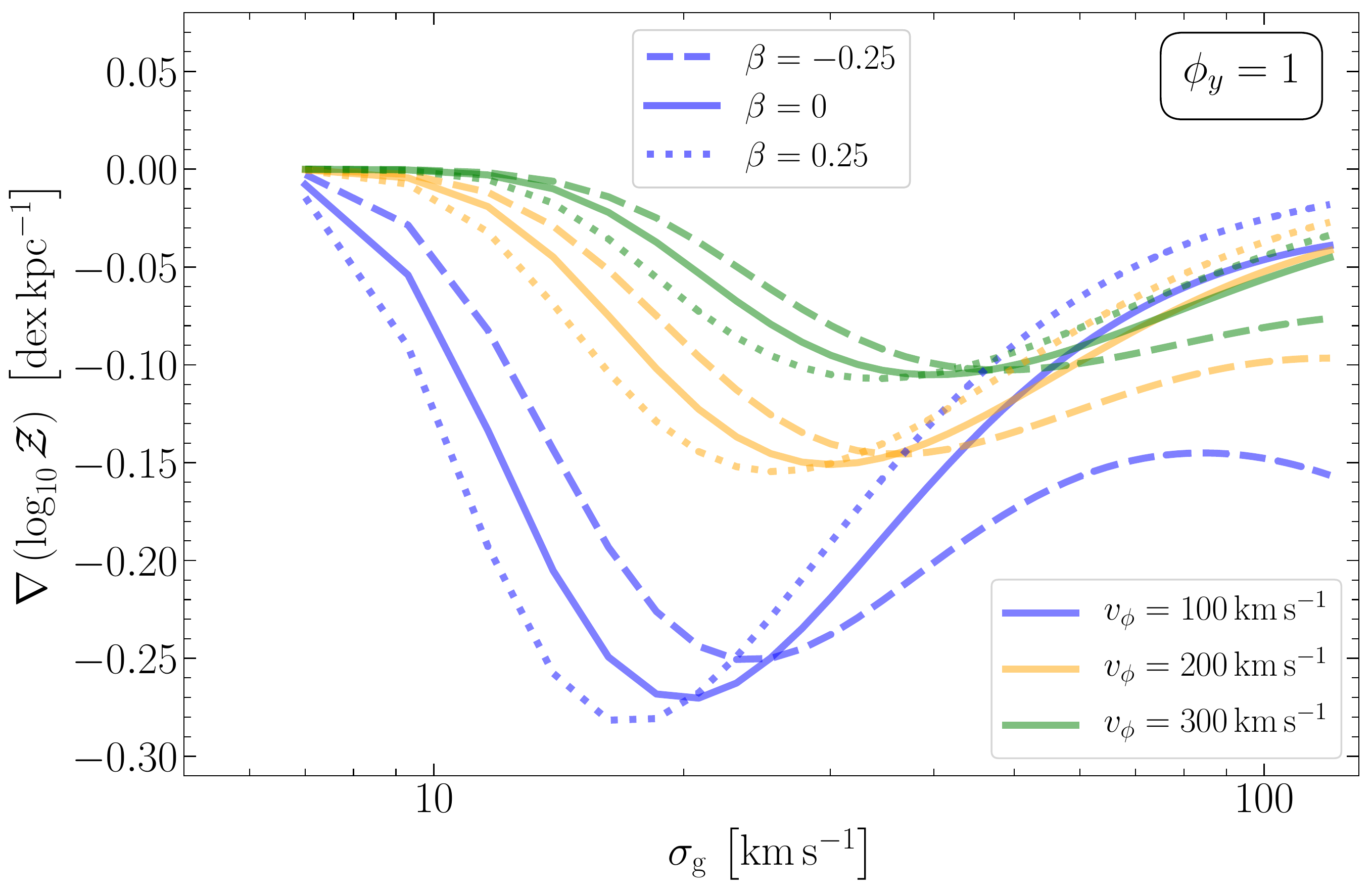}
\caption{Metallicity gradients from the model for different values of $v_{\phi}$ and the rotation curve index $\beta$ at fixed $z=2$. The curves are only plotted for the highest possible yield reduction factor $\phi_y=1$, thus providing a limit on the most negative metallicity gradient the model can produce given a set of input parameters. The main takeaway from this plot is that high-$z$ galaxies that are very turbulent (high $\sigma_{\rm{g}}$) but show falling rotation curves ($\beta<0$) can still maintain a steep metallicity gradient in equilibrium.}
\label{fig:grad_sigma_beta}
\end{figure}

\autoref{fig:grad_sigma_beta} shows the same model curves as in \autoref{fig:grad_sigma_model}, but with three different rotation curve indices, $\beta=-0.25,\,0$, and $0.25$. For the sake of clarity, we do not overplot the observational data in this figure. While changing $\beta$ does not significantly change the range of metallicity gradients produced by the model for large $v_{\phi}$ (\textit{i.e.,} more massive galaxies), it has some effect for galaxies with smaller $v_{\phi}$. If $\beta < 0$, the model allows for steeper gradients (by a factor of 3) for low-mass galaxies with high $\sigma_{\rm{g}}$. This is because as compared to the default $\beta=0$, the Péclet number $\mathcal{P}$ decreases by a larger factor than the source term $\mathcal{S}$ when $\beta<0$ \citep[equations~38 and~39]{2020aMNRAS.xxx..xxxS}. Thus, $\mathcal{S}$ dominates, giving rise to steeper gradients. On the other hand, $\mathcal{P}$ increases by a larger factor than $\mathcal{S}$ for $\beta > 0$ as compared to the default $\beta=0$. Thus, $\mathcal{P}$ dominates, giving rise to flatter gradients. This analysis tells us that high-$z$ galaxies with high levels of turbulence and falling rotation curves ($\beta<0$) can still maintain a steep metallicity gradient due to the decreased strength of advection as compared to metal production.

With respect to $\beta$, a detailed comparison of the model with observational data is beyond the scope of the present study. Future observations will provide further constraints to the model parameters such as varying $\beta$ and $Q$.

\section{Conclusions}
\label{s:conclusion}
In this work, we explore the relationship between gas kinematics (rotational velocity $v_{\phi}$ and velocity dispersion $\sigma_{\rm{g}}$) and gas phase metallicity gradients in star-forming galaxies at $0.1 \leq z \leq 2.5$ using a compilation of 74 galaxies across 5 ground based IFU spectroscopy samples, and our new analytic model \citep{2020aMNRAS.xxx..xxxS}. To partially alleviate the inhomogeneities in the compiled data that used diverse instruments and techniques, we reanalyse the kinematics for all galaxies following \cite{2018ApJS..238...21F}. All the samples (except for one) use the [\ion{N}{ii}]/H$\alpha$ ratio and the \cite{2004MNRAS.348L..59P} calibration to obtain the metallicity gradients.

We find that high-redshift galaxies that are highly turbulent ($\sigma_{\rm{g}} > 60\,\mathrm{km\,s^{-1}}$) show shallow or flat metallicity gradients ($> -0.05\,\rm{dex\,kpc^{-1}}$), whereas galaxies with intermediate levels of turbulence ($\sigma_{\rm{g}} \approx 20 - 50\,\mathrm{km\,s^{-1}}$) show comparatively the largest scatter in their measured metallicity gradients. Finally, galaxies with the lowest $\sigma_{\rm{g}}$ ($< 20\,\mathrm{km\,s^{-1}}$) show flat gradients, although the small number of low-$\sigma_{\rm{g}}$ galaxies in our sample renders this conclusion tentative. Our findings are consistent with the predictions made by simulations of galaxy formation (FIRE and IllustrisTNG50), which find that steep negative metallicity gradients only occur in galaxies that are rotation-dominated ($v_{\phi}/\sigma_{\rm{g}} > 1$), whereas all dispersion-dominated galaxies show relatively flat gradients \citep{2017MNRAS.466.4780M,2020arXiv200710993H}.

We compare the data against predictions from our recently developed model of gas phase metallicity gradients in galaxies \citep{2020aMNRAS.xxx..xxxS} to provide a physical explanation for the observed trends. We find that the model is able to reproduce the observed, non-monotonic relationship between metallicity gradients and gas kinematics. Strong inward advection of gas leads to efficient metal mixing when the gas velocity dispersion is high. This mixing results in flatter gradients. However, the relationship between velocity dispersion and inward advection rate also depends on the index of the galaxy rotation curve -- galaxies with falling rotation curves can maintain high velocity dispersion with relatively lower inflow rates, and thus can retain steeper metal gradients than their counterparts with flat or rising rotation curves. In contrast, the flat gradients seen with low gas velocity dispersion are due to stronger cosmic accretion of metal-poor gas which dilutes the central regions of galaxies. In these cases of high and low gas velocity dispersion, advection and accretion respectively dominate over metal production which is otherwise responsible for creating negative gradients which follow the star formation profile. 

The steepest gradients as well as the largest scatter in the gradients in the model are found for intermediate velocity dispersions where both the inward advection of gas and cosmic accretion of metal-poor gas are weak compared to metal production. The scatter at intermediate velocity dispersions may arise from galaxy-to-galaxy variations in the preferential ejection of metals through galactic winds before they mix with the ISM: the most negative metallicity gradients arise in galaxies where metals mix efficiently with the ISM before ejection, while flatter gradients occur in galaxies where a substantial fraction of supernova-produced metals are ejected directly into galactic winds before mixing with the ISM. However, we note the large number of observational uncertainties which may also dominate this scatter. We also find that while metal diffusion is also an important process that contributes to flattening the metallicity gradients, it never simultaneously dominates inward advection and cosmic accretion.

While our metallicity evolution model successfully reproduces the observed shape and scatter of the non-linear relationship between metallicity gradients and gas kinematics in high-redshift galaxies, it is in better agreement with the data in the gradient$-\sigma_{\rm{g}}$ space than that in the gradient$-v_{\phi}/\sigma_{\rm{g}}$ space. The model also cannot predict the full distribution of galaxies in either set of parameters without better constraints on the metal enrichment of outflows. The current sample only consists of 74~galaxies, and there is clearly scope for more observations of galaxies at high redshift against which we can test our inferences about the physics behind the impacts of gas kinematics on metallicity gradients.

\section*{Acknowledgements}
\addcontentsline{toc}{section}{Acknowledgements}
We thank the anonymous reviewer for a constructive feedback on the manuscript. We also thank John Forbes for going through a preprint of this paper and providing comments. We are grateful to Xiangcheng Ma, Ben\^oit Epinat, Natascha Förster-Schreiber, Chao-Ling Hung, and Henry Poetrodjojo for kindly sharing their simulation and/or observational data, and to Nicha Leethochawalit, Lisa Kewley, Ayan Acharyya, Brent Groves and Philipp Lang for useful discussions. PS is supported by the Australian Government Research Training Program (RTP) Scholarship. PS and EW acknowledge support by the Australian Research Council Centre of Excellence for All Sky Astrophysics in 3 Dimensions (ASTRO 3D), through project number CE170100013. MRK and CF acknowledge funding provided by the Australian Research Council (ARC) through Discovery Projects DP190101258 (MRK) and DP170100603 (CF) and Future Fellowships FT180100375 (MRK) and FT180100495 (CF). MRK is also the recipient of an Alexander von Humboldt award. CF further acknowledges an Australia-Germany Joint Research Cooperation Scheme grant (UA-DAAD). Parts of this paper were written during the ASTRO 3D writing retreat 2019. This work is based on observations taken by the MUSE-Wide Survey as part of the MUSE Consortium \citep{2017A&A...606A..12H,2019A&A...624A.141U}. This work is also based on observations collected at the European Southern Observatory under ESO programmes 60.A-9460, 092.A-0091, 093.A-0079 and 094.A-0205, 179.A-0823, 75.A-0318, 78.A-0177, and 084.B-0300. This research has made extensive use of Python packages \verb'astropy' \citep{2013A&A...558A..33A,2018AJ....156..123A}, \verb'numpy' \citep{oliphant2006guide,2020arXiv200610256H}, \verb'scipy' \citep{2020NatMe..17..261V} and \verb'Matplotlib' \citep{Hunter:2007}. This research has also made extensive use of NASA's Astrophysics Data System Bibliographic Services, the \cite{2006PASP..118.1711W} cosmology calculator, and the image-to-data tool \texttt{WebPlotDigitizer}.

\section*{Data availability statement}
No new data were generated for this work. The compiled data with reanalysed kinematics is available in \aref{s:app_data}.




\bibliographystyle{mnras}
\bibliography{references} 

\begin{thebibliography}{}
\makeatletter
\relax
\def\mn@urlcharsother{\let\do\@makeother \do\$\do\&\do\#\do\^\do\_\do\%\do\~}
\def\mn@doi{\begingroup\mn@urlcharsother \@ifnextchar [ {\mn@doi@}
  {\mn@doi@[]}}
\def\mn@doi@[#1]#2{\def\@tempa{#1}\ifx\@tempa\@empty \href
  {http://dx.doi.org/#2} {doi:#2}\else \href {http://dx.doi.org/#2} {#1}\fi
  \endgroup}
\def\mn@eprint#1#2{\mn@eprint@#1:#2::\@nil}
\def\mn@eprint@arXiv#1{\href {http://arxiv.org/abs/#1} {{\tt arXiv:#1}}}
\def\mn@eprint@dblp#1{\href {http://dblp.uni-trier.de/rec/bibtex/#1.xml}
  {dblp:#1}}
\def\mn@eprint@#1:#2:#3:#4\@nil{\def\@tempa {#1}\def\@tempb {#2}\def\@tempc
  {#3}\ifx \@tempc \@empty \let \@tempc \@tempb \let \@tempb \@tempa \fi \ifx
  \@tempb \@empty \def\@tempb {arXiv}\fi \@ifundefined
  {mn@eprint@\@tempb}{\@tempb:\@tempc}{\expandafter \expandafter \csname
  mn@eprint@\@tempb\endcsname \expandafter{\@tempc}}}

\bibitem[\protect\citeauthoryear{{Aller}}{{Aller}}{1942}]{1942ApJ....95...52A}
{Aller} L.~H.,  1942, \mn@doi [\apj] {10.1086/144372}, \href
  {https://ui.adsabs.harvard.edu/abs/1942ApJ....95...52A} {95, 52}

\bibitem[\protect\citeauthoryear{{Armillotta}, {Krumholz}  \&
  {Fujimoto}}{{Armillotta} et~al.}{2018}]{2018MNRAS.481.5000A}
{Armillotta} L.,  {Krumholz} M.~R.,   {Fujimoto} Y.,  2018, \mn@doi [\mnras]
  {10.1093/mnras/sty2625}, \href
  {https://ui.adsabs.harvard.edu/abs/2018MNRAS.481.5000A} {481, 5000}

\bibitem[\protect\citeauthoryear{{Asplund}, {Grevesse}, {Sauval}  \&
  {Scott}}{{Asplund} et~al.}{2009}]{2009ARA&A..47..481A}
{Asplund} M.,  {Grevesse} N.,  {Sauval} A.~J.,   {Scott} P.,  2009, \mn@doi
  [\araa] {10.1146/annurev.astro.46.060407.145222}, \href
  {https://ui.adsabs.harvard.edu/abs/2009ARA&A..47..481A} {47, 481}

\bibitem[\protect\citeauthoryear{{Astropy Collaboration} et~al.,}{{Astropy
  Collaboration} et~al.}{2013}]{2013A&A...558A..33A}
{Astropy Collaboration} et~al., 2013, \mn@doi [\aap]
  {10.1051/0004-6361/201322068}, \href
  {https://ui.adsabs.harvard.edu/abs/2013A\%26A...558A..33A} {558, A33}

\bibitem[\protect\citeauthoryear{{Astropy Collaboration} et~al.,}{{Astropy
  Collaboration} et~al.}{2018}]{2018AJ....156..123A}
{Astropy Collaboration} et~al., 2018, \mn@doi [\aj] {10.3847/1538-3881/aabc4f},
  \href {https://ui.adsabs.harvard.edu/abs/2018AJ....156..123A} {156, 123}

\bibitem[\protect\citeauthoryear{{Bacon} et~al.,}{{Bacon}
  et~al.}{2010}]{2010SPIE.7735E..08B}
{Bacon} R.,  et~al., 2010, in \procspie. p. 773508, \mn@doi{10.1117/12.856027}

\bibitem[\protect\citeauthoryear{{Baldwin}, {Phillips}  \&
  {Terlevich}}{{Baldwin} et~al.}{1981}]{1981PASP...93....5B}
{Baldwin} J.~A.,  {Phillips} M.~M.,   {Terlevich} R.,  1981, \mn@doi [\pasp]
  {10.1086/130766}, \href
  {https://ui.adsabs.harvard.edu/abs/1981PASP...93....5B} {93, 5}

\bibitem[\protect\citeauthoryear{{Belfiore} et~al.,}{{Belfiore}
  et~al.}{2017}]{2017MNRAS.469..151B}
{Belfiore} F.,  et~al., 2017, \mn@doi [\mnras] {10.1093/mnras/stx789}, \href
  {https://ui.adsabs.harvard.edu/abs/2017MNRAS.469..151B} {469, 151}

\bibitem[\protect\citeauthoryear{{Bell} \& {de Jong}}{{Bell} \& {de
  Jong}}{2001}]{2001ApJ...550..212B}
{Bell} E.~F.,  {de Jong} R.~S.,  2001, \mn@doi [\apj] {10.1086/319728}, \href
  {https://ui.adsabs.harvard.edu/abs/2001ApJ...550..212B} {550, 212}

\bibitem[\protect\citeauthoryear{{Bonnet} et~al.,}{{Bonnet}
  et~al.}{2004}]{2004SPIE.5490..130B}
{Bonnet} H.,  et~al., 2004, in {Bonaccini Calia} D.,  {Ellerbroek} B.~L.,
  {Ragazzoni} R.,  eds,  Society of Photo-Optical Instrumentation Engineers
  (SPIE) Conference Series Vol. 5490, \procspie. pp 130--138,
  \mn@doi{10.1117/12.551187}

\bibitem[\protect\citeauthoryear{{Burkert} et~al.,}{{Burkert}
  et~al.}{2010}]{2010ApJ...725.2324B}
{Burkert} A.,  et~al., 2010, \mn@doi [\apj] {10.1088/0004-637X/725/2/2324},
  \href {https://ui.adsabs.harvard.edu/abs/2010ApJ...725.2324B} {725, 2324}

\bibitem[\protect\citeauthoryear{{Burkert} et~al.,}{{Burkert}
  et~al.}{2016}]{2016ApJ...826..214B}
{Burkert} A.,  et~al., 2016, \mn@doi [\apj] {10.3847/0004-637X/826/2/214},
  \href {https://ui.adsabs.harvard.edu/abs/2016ApJ...826..214B} {826, 214}

\bibitem[\protect\citeauthoryear{{Calzetti}}{{Calzetti}}{2001}]{2001PASP..113.1449C}
{Calzetti} D.,  2001, \mn@doi [\pasp] {10.1086/324269}, \href
  {https://ui.adsabs.harvard.edu/abs/2001PASP..113.1449C} {113, 1449}

\bibitem[\protect\citeauthoryear{{Carton} et~al.,}{{Carton}
  et~al.}{2017}]{2017MNRAS.468.2140C}
{Carton} D.,  et~al., 2017, \mn@doi [\mnras] {10.1093/mnras/stx545}, \href
  {https://ui.adsabs.harvard.edu/abs/2017MNRAS.468.2140C} {468, 2140}

\bibitem[\protect\citeauthoryear{{Carton} et~al.,}{{Carton}
  et~al.}{2018}]{2018MNRAS.478.4293C}
{Carton} D.,  et~al., 2018, \mn@doi [\mnras] {10.1093/mnras/sty1343}, \href
  {https://ui.adsabs.harvard.edu/abs/2018MNRAS.478.4293C} {478, 4293}

\bibitem[\protect\citeauthoryear{{Chabrier}}{{Chabrier}}{2003}]{2003PASP..115..763C}
{Chabrier} G.,  2003, \mn@doi [\pasp] {10.1086/376392}, \href
  {https://ui.adsabs.harvard.edu/abs/2003PASP..115..763C} {115, 763}

\bibitem[\protect\citeauthoryear{{Contini} et~al.,}{{Contini}
  et~al.}{2012}]{2012A&A...539A..91C}
{Contini} T.,  et~al., 2012, \mn@doi [\aap] {10.1051/0004-6361/201117541},
  \href {https://ui.adsabs.harvard.edu/abs/2012A&A...539A..91C} {539, A91}

\bibitem[\protect\citeauthoryear{{Curti} et~al.,}{{Curti}
  et~al.}{2020}]{2020MNRAS.492..821C}
{Curti} M.,  et~al., 2020, \mn@doi [\mnras] {10.1093/mnras/stz3379}, \href
  {https://ui.adsabs.harvard.edu/abs/2020MNRAS.492..821C} {492, 821}

\bibitem[\protect\citeauthoryear{{Davies} et~al.,}{{Davies}
  et~al.}{1997}]{1997SPIE.2871.1099D}
{Davies} R.~L.,  et~al., 1997, in {Ardeberg} A.~L.,  ed.,  Society of
  Photo-Optical Instrumentation Engineers (SPIE) Conference Series Vol. 2871,
  Optical Telescopes of Today and Tomorrow. pp 1099--1106,
  \mn@doi{10.1117/12.268996}

\bibitem[\protect\citeauthoryear{{Davies} et~al.,}{{Davies}
  et~al.}{2011}]{2011ApJ...741...69D}
{Davies} R.,  et~al., 2011, \mn@doi [\apj] {10.1088/0004-637X/741/2/69}, \href
  {https://ui.adsabs.harvard.edu/abs/2011ApJ...741...69D} {741, 69}

\bibitem[\protect\citeauthoryear{{Davies} et~al.,}{{Davies}
  et~al.}{2018}]{2018SPIE10702E..09D}
{Davies} R.,  et~al., 2018, in {Evans} C.~J.,  {Simard} L.,   {Takami} H.,
  eds,  Society of Photo-Optical Instrumentation Engineers (SPIE) Conference
  Series Vol. 10702, Ground-based and Airborne Instrumentation for Astronomy
  VII. p. 1070209 (\mn@eprint {arXiv} {1807.05089}),
  \mn@doi{10.1117/12.2311480}

\bibitem[\protect\citeauthoryear{{Davies} et~al.,}{{Davies}
  et~al.}{2021}]{2020arXiv201210445D}
{Davies} R.~L.,  et~al., 2021, \mn@doi [\apj] {10.3847/1538-4357/abd551}, \href
  {https://ui.adsabs.harvard.edu/abs/2021ApJ...909...78D} {909, 78}

\bibitem[\protect\citeauthoryear{{Di Teodoro} \& {Fraternali}}{{Di Teodoro} \&
  {Fraternali}}{2015}]{2015MNRAS.451.3021D}
{Di Teodoro} E.~M.,  {Fraternali} F.,  2015, \mn@doi [\mnras]
  {10.1093/mnras/stv1213}, \href
  {https://ui.adsabs.harvard.edu/abs/2015MNRAS.451.3021D} {451, 3021}

\bibitem[\protect\citeauthoryear{{Dutton} et~al.,}{{Dutton}
  et~al.}{2011}]{2011MNRAS.410.1660D}
{Dutton} A.~A.,  et~al., 2011, \mn@doi [\mnras]
  {10.1111/j.1365-2966.2010.17555.x}, \href
  {https://ui.adsabs.harvard.edu/abs/2011MNRAS.410.1660D} {410, 1660}

\bibitem[\protect\citeauthoryear{{Eisenhauer} et~al.,}{{Eisenhauer}
  et~al.}{2003}]{2003SPIE.4841.1548E}
{Eisenhauer} F.,  et~al., 2003, in {Iye} M.,  {Moorwood} A. F.~M.,  eds,
  Society of Photo-Optical Instrumentation Engineers (SPIE) Conference Series
  Vol. 4841, \procspie. pp 1548--1561 (\mn@eprint {arXiv} {astro-ph/0306191}),
  \mn@doi{10.1117/12.459468}

\bibitem[\protect\citeauthoryear{{Ellis} et~al.,}{{Ellis}
  et~al.}{2020}]{2020SPIE11447E..A0E}
{Ellis} S.,  et~al., 2020, in Society of Photo-Optical Instrumentation
  Engineers (SPIE) Conference Series. p. 11447A0, \mn@doi{10.1117/12.2561930}

\bibitem[\protect\citeauthoryear{{Epinat} et~al.,}{{Epinat}
  et~al.}{2012}]{2012A&A...539A..92E}
{Epinat} B.,  et~al., 2012, \mn@doi [\aap] {10.1051/0004-6361/201117711}, \href
  {https://ui.adsabs.harvard.edu/abs/2012A&A...539A..92E} {539, A92}

\bibitem[\protect\citeauthoryear{{Faucher-Gigu{\`e}re}, {Kere{\v{s}}}  \&
  {Ma}}{{Faucher-Gigu{\`e}re} et~al.}{2011}]{2011MNRAS.417.2982F}
{Faucher-Gigu{\`e}re} C.-A.,  {Kere{\v{s}}} D.,   {Ma} C.-P.,  2011, \mn@doi
  [\mnras] {10.1111/j.1365-2966.2011.19457.x}, \href
  {https://ui.adsabs.harvard.edu/abs/2011MNRAS.417.2982F} {417, 2982}

\bibitem[\protect\citeauthoryear{{Faucher-Gigu{\`e}re}, {Quataert}  \&
  {Hopkins}}{{Faucher-Gigu{\`e}re} et~al.}{2013}]{2013MNRAS.433.1970F}
{Faucher-Gigu{\`e}re} C.-A.,  {Quataert} E.,   {Hopkins} P.~F.,  2013, \mn@doi
  [\mnras] {10.1093/mnras/stt866}, \href
  {https://ui.adsabs.harvard.edu/abs/2013MNRAS.433.1970F} {433, 1970}

\bibitem[\protect\citeauthoryear{{Fern{\'a}ndez-Ontiveros}
  et~al.,}{{Fern{\'a}ndez-Ontiveros} et~al.}{2017}]{2017PASA...34...53F}
{Fern{\'a}ndez-Ontiveros} J.~A.,  et~al., 2017, \mn@doi [\pasa]
  {10.1017/pasa.2017.43}, \href
  {https://ui.adsabs.harvard.edu/abs/2017PASA...34...53F} {34, e053}

\bibitem[\protect\citeauthoryear{{Forbes}, {Krumholz}, {Burkert}  \&
  {Dekel}}{{Forbes} et~al.}{2014}]{2014MNRAS.438.1552F}
{Forbes} J.~C.,  {Krumholz} M.~R.,  {Burkert} A.,   {Dekel} A.,  2014, \mn@doi
  [\mnras] {10.1093/mnras/stt2294}, \href
  {https://ui.adsabs.harvard.edu/abs/2014MNRAS.438.1552F} {438, 1552}

\bibitem[\protect\citeauthoryear{{F{\"o}rster Schreiber} \&
  {Wuyts}}{{F{\"o}rster Schreiber} \& {Wuyts}}{2020}]{2020ARA&A..58..661F}
{F{\"o}rster Schreiber} N.~M.,  {Wuyts} S.,  2020, \mn@doi [\araa]
  {10.1146/annurev-astro-032620-021910}, \href
  {https://ui.adsabs.harvard.edu/abs/2020ARA&A..58..661F} {58, 661}

\bibitem[\protect\citeauthoryear{{F{\"o}rster Schreiber} et~al.,}{{F{\"o}rster
  Schreiber} et~al.}{2009}]{2009ApJ...706.1364F}
{F{\"o}rster Schreiber} N.~M.,  et~al., 2009, \mn@doi [\apj]
  {10.1088/0004-637X/706/2/1364}, \href
  {https://ui.adsabs.harvard.edu/abs/2009ApJ...706.1364F} {706, 1364}

\bibitem[\protect\citeauthoryear{{F{\"o}rster Schreiber} et~al.,}{{F{\"o}rster
  Schreiber} et~al.}{2018}]{2018ApJS..238...21F}
{F{\"o}rster Schreiber} N.~M.,  et~al., 2018, \mn@doi [\apjs]
  {10.3847/1538-4365/aadd49}, \href
  {https://ui.adsabs.harvard.edu/abs/2018ApJS..238...21F} {238, 21}

\bibitem[\protect\citeauthoryear{{Foster} et~al.,}{{Foster}
  et~al.}{2020}]{2020arXiv201113567F}
{Foster} C.,  et~al., 2020, arXiv e-prints, \href
  {https://ui.adsabs.harvard.edu/abs/2020arXiv201113567F} {p. arXiv:2011.13567}

\bibitem[\protect\citeauthoryear{{Furlanetto}}{{Furlanetto}}{2021}]{2021MNRAS.500.3394F}
{Furlanetto} S.~R.,  2021, \mn@doi [\mnras] {10.1093/mnras/staa3451}, \href
  {https://ui.adsabs.harvard.edu/abs/2021MNRAS.500.3394F} {500, 3394}

\bibitem[\protect\citeauthoryear{{Garnett}}{{Garnett}}{2002}]{2002ApJ...581.1019G}
{Garnett} D.~R.,  2002, \mn@doi [\apj] {10.1086/344301}, \href
  {https://ui.adsabs.harvard.edu/abs/2002ApJ...581.1019G} {581, 1019}

\bibitem[\protect\citeauthoryear{{Gentry}, {Krumholz}, {Madau}  \&
  {Lupi}}{{Gentry} et~al.}{2019}]{Gentry19a}
{Gentry} E.~S.,  {Krumholz} M.~R.,  {Madau} P.,   {Lupi} A.,  2019, \mn@doi
  [\mnras] {10.1093/mnras/sty3319}, \href
  {https://ui.adsabs.harvard.edu/\#abs/2019MNRAS.483.3647G} {483, 3647}

\bibitem[\protect\citeauthoryear{{Genzel} et~al.,}{{Genzel}
  et~al.}{2017}]{2017Natur.543..397G}
{Genzel} R.,  et~al., 2017, \mn@doi [\nat] {10.1038/nature21685}, \href
  {https://ui.adsabs.harvard.edu/abs/2017Natur.543..397G} {543, 397}

\bibitem[\protect\citeauthoryear{{Genzel} et~al.,}{{Genzel}
  et~al.}{2020}]{2020ApJ...902...98G}
{Genzel} R.,  et~al., 2020, \mn@doi [\apj] {10.3847/1538-4357/abb0ea}, \href
  {https://ui.adsabs.harvard.edu/abs/2020ApJ...902...98G} {902, 98}

\bibitem[\protect\citeauthoryear{{Gibson}, {Pilkington}, {Brook}, {Stinson}  \&
  {Bailin}}{{Gibson} et~al.}{2013}]{2013A&A...554A..47G}
{Gibson} B.~K.,  {Pilkington} K.,  {Brook} C.~B.,  {Stinson} G.~S.,   {Bailin}
  J.,  2013, \mn@doi [\aap] {10.1051/0004-6361/201321239}, \href
  {https://ui.adsabs.harvard.edu/abs/2013A&A...554A..47G} {554, A47}

\bibitem[\protect\citeauthoryear{{Gillman} et~al.,}{{Gillman}
  et~al.}{2021}]{2021MNRAS.500.4229G}
{Gillman} S.,  et~al., 2021, \mn@doi [\mnras] {10.1093/mnras/staa3400}, \href
  {https://ui.adsabs.harvard.edu/abs/2021MNRAS.500.4229G} {500, 4229}

\bibitem[\protect\citeauthoryear{{Glowacki}, {Elson}  \& {Dav{\'e}}}{{Glowacki}
  et~al.}{2020}]{2020arXiv201108866G}
{Glowacki} M.,  {Elson} E.,   {Dav{\'e}} R.,  2020, arXiv e-prints, \href
  {https://ui.adsabs.harvard.edu/abs/2020arXiv201108866G} {p. arXiv:2011.08866}

\bibitem[\protect\citeauthoryear{{Goldbaum}, {Krumholz}  \&
  {Forbes}}{{Goldbaum} et~al.}{2016}]{2016ApJ...827...28G}
{Goldbaum} N.~J.,  {Krumholz} M.~R.,   {Forbes} J.~C.,  2016, \mn@doi [\apj]
  {10.3847/0004-637X/827/1/28}, \href
  {https://ui.adsabs.harvard.edu/abs/2016ApJ...827...28G} {827, 28}

\bibitem[\protect\citeauthoryear{{Grogin} et~al.,}{{Grogin}
  et~al.}{2011}]{2011ApJS..197...35G}
{Grogin} N.~A.,  et~al., 2011, \mn@doi [\apjs] {10.1088/0067-0049/197/2/35},
  \href {https://ui.adsabs.harvard.edu/abs/2011ApJS..197...35G} {197, 35}

\bibitem[\protect\citeauthoryear{{Harris} et~al.,}{{Harris}
  et~al.}{2020}]{2020arXiv200610256H}
{Harris} C.~R.,  et~al., 2020, \mn@doi [\nat] {10.1038/s41586-020-2649-2},
  \href {https://ui.adsabs.harvard.edu/abs/2020arXiv200610256H} {585, 357}

\bibitem[\protect\citeauthoryear{{Hemler} et~al.,}{{Hemler}
  et~al.}{2020}]{2020arXiv200710993H}
{Hemler} Z.~S.,  et~al., 2020, arXiv e-prints, \href
  {https://ui.adsabs.harvard.edu/abs/2020arXiv200710993H} {p. arXiv:2007.10993}

\bibitem[\protect\citeauthoryear{{Herenz} et~al.,}{{Herenz}
  et~al.}{2017}]{2017A&A...606A..12H}
{Herenz} E.~C.,  et~al., 2017, \mn@doi [\aap] {10.1051/0004-6361/201731055},
  \href {https://ui.adsabs.harvard.edu/abs/2017A&A...606A..12H} {606, A12}

\bibitem[\protect\citeauthoryear{{Ho} et~al.,}{{Ho}
  et~al.}{2016}]{2016Ap&SS.361..280H}
{Ho} I.~T.,  et~al., 2016, \mn@doi [\apss] {10.1007/s10509-016-2865-2}, \href
  {https://ui.adsabs.harvard.edu/abs/2016Ap&SS.361..280H} {361, 280}

\bibitem[\protect\citeauthoryear{{Hopkins}, {Kere{\v{s}}}, {O{\~n}orbe},
  {Faucher-Gigu{\`e}re}, {Quataert}, {Murray}  \& {Bullock}}{{Hopkins}
  et~al.}{2014}]{2014MNRAS.445..581H}
{Hopkins} P.~F.,  {Kere{\v{s}}} D.,  {O{\~n}orbe} J.,  {Faucher-Gigu{\`e}re}
  C.-A.,  {Quataert} E.,  {Murray} N.,   {Bullock} J.~S.,  2014, \mn@doi
  [\mnras] {10.1093/mnras/stu1738}, \href
  {https://ui.adsabs.harvard.edu/abs/2014MNRAS.445..581H} {445, 581}

\bibitem[\protect\citeauthoryear{{Hopkins} et~al.,}{{Hopkins}
  et~al.}{2018}]{2018MNRAS.480..800H}
{Hopkins} P.~F.,  et~al., 2018, \mn@doi [\mnras] {10.1093/mnras/sty1690}, \href
  {https://ui.adsabs.harvard.edu/abs/2018MNRAS.480..800H} {480, 800}

\bibitem[\protect\citeauthoryear{Hunter}{Hunter}{2007}]{Hunter:2007}
Hunter J.~D.,  2007, \mn@doi [Computing in Science \& Engineering]
  {10.1109/MCSE.2007.55}, 9, 90

\bibitem[\protect\citeauthoryear{{Jones}, {Ellis}, {Richard}  \&
  {Jullo}}{{Jones} et~al.}{2013}]{2013ApJ...765...48J}
{Jones} T.,  {Ellis} R.~S.,  {Richard} J.,   {Jullo} E.,  2013, \mn@doi [\apj]
  {10.1088/0004-637X/765/1/48}, \href
  {https://ui.adsabs.harvard.edu/abs/2013ApJ...765...48J} {765, 48}

\bibitem[\protect\citeauthoryear{{Jones} et~al.,}{{Jones}
  et~al.}{2015}]{2015AJ....149..107J}
{Jones} T.,  et~al., 2015, \mn@doi [\aj] {10.1088/0004-6256/149/3/107}, \href
  {https://ui.adsabs.harvard.edu/abs/2015AJ....149..107J} {149, 107}

\bibitem[\protect\citeauthoryear{{Kassin} et~al.,}{{Kassin}
  et~al.}{2012}]{2012ApJ...758..106K}
{Kassin} S.~A.,  et~al., 2012, \mn@doi [\apj] {10.1088/0004-637X/758/2/106},
  \href {https://ui.adsabs.harvard.edu/abs/2012ApJ...758..106K} {758, 106}

\bibitem[\protect\citeauthoryear{{Kassin}, {Brooks}, {Governato}, {Weiner}  \&
  {Gardner}}{{Kassin} et~al.}{2014}]{2014ApJ...790...89K}
{Kassin} S.~A.,  {Brooks} A.,  {Governato} F.,  {Weiner} B.~J.,   {Gardner}
  J.~P.,  2014, \mn@doi [\apj] {10.1088/0004-637X/790/2/89}, \href
  {https://ui.adsabs.harvard.edu/abs/2014ApJ...790...89K} {790, 89}

\bibitem[\protect\citeauthoryear{{Kennicutt} \& {Evans}}{{Kennicutt} \&
  {Evans}}{2012}]{2012ARA&A..50..531K}
{Kennicutt} R.~C.,  {Evans} N.~J.,  2012, \mn@doi [\araa]
  {10.1146/annurev-astro-081811-125610}, \href
  {https://ui.adsabs.harvard.edu/abs/2012ARA&A..50..531K} {50, 531}

\bibitem[\protect\citeauthoryear{{Kewley} \& {Ellison}}{{Kewley} \&
  {Ellison}}{2008}]{2008ApJ...681.1183K}
{Kewley} L.~J.,  {Ellison} S.~L.,  2008, \mn@doi [\apj] {10.1086/587500}, \href
  {https://ui.adsabs.harvard.edu/abs/2008ApJ...681.1183K} {681, 1183}

\bibitem[\protect\citeauthoryear{{Kewley}, {Heisler}, {Dopita}  \&
  {Lumsden}}{{Kewley} et~al.}{2001}]{2001ApJS..132...37K}
{Kewley} L.~J.,  {Heisler} C.~A.,  {Dopita} M.~A.,   {Lumsden} S.,  2001,
  \mn@doi [\apjs] {10.1086/318944}, \href
  {https://ui.adsabs.harvard.edu/abs/2001ApJS..132...37K} {132, 37}

\bibitem[\protect\citeauthoryear{{Kewley}, {Groves}, {Kauffmann}  \&
  {Heckman}}{{Kewley} et~al.}{2006}]{2006MNRAS.372..961K}
{Kewley} L.~J.,  {Groves} B.,  {Kauffmann} G.,   {Heckman} T.,  2006, \mn@doi
  [\mnras] {10.1111/j.1365-2966.2006.10859.x}, \href
  {https://ui.adsabs.harvard.edu/abs/2006MNRAS.372..961K} {372, 961}

\bibitem[\protect\citeauthoryear{{Kewley}, {Maier}, {Yabe}, {Ohta}, {Akiyama},
  {Dopita}  \& {Yuan}}{{Kewley} et~al.}{2013a}]{2013ApJ...774L..10K}
{Kewley} L.~J.,  {Maier} C.,  {Yabe} K.,  {Ohta} K.,  {Akiyama} M.,  {Dopita}
  M.~A.,   {Yuan} T.,  2013a, \mn@doi [\apjl] {10.1088/2041-8205/774/1/L10},
  \href {https://ui.adsabs.harvard.edu/abs/2013ApJ...774L..10K} {774, L10}

\bibitem[\protect\citeauthoryear{{Kewley}, {Dopita}, {Leitherer}, {Dav{\'e}},
  {Yuan}, {Allen}, {Groves}  \& {Sutherland}}{{Kewley}
  et~al.}{2013b}]{2013ApJ...774..100K}
{Kewley} L.~J.,  {Dopita} M.~A.,  {Leitherer} C.,  {Dav{\'e}} R.,  {Yuan} T.,
  {Allen} M.,  {Groves} B.,   {Sutherland} R.,  2013b, \mn@doi [\apj]
  {10.1088/0004-637X/774/2/100}, \href
  {https://ui.adsabs.harvard.edu/abs/2013ApJ...774..100K} {774, 100}

\bibitem[\protect\citeauthoryear{{Kewley}, {Nicholls}  \&
  {Sutherland}}{{Kewley} et~al.}{2019}]{2019ARA&A..57..511K}
{Kewley} L.~J.,  {Nicholls} D.~C.,   {Sutherland} R.~S.,  2019, \mn@doi [\araa]
  {10.1146/annurev-astro-081817-051832}, \href
  {https://ui.adsabs.harvard.edu/abs/2019ARA&A..57..511K} {57, 511}

\bibitem[\protect\citeauthoryear{{Kim} \& {Ostriker}}{{Kim} \&
  {Ostriker}}{2015}]{2015ApJ...815...67K}
{Kim} C.-G.,  {Ostriker} E.~C.,  2015, \mn@doi [\apj]
  {10.1088/0004-637X/815/1/67}, \href
  {https://ui.adsabs.harvard.edu/abs/2015ApJ...815...67K} {815, 67}

\bibitem[\protect\citeauthoryear{{Koekemoer} et~al.,}{{Koekemoer}
  et~al.}{2011}]{2011ApJS..197...36K}
{Koekemoer} A.~M.,  et~al., 2011, \mn@doi [\apjs] {10.1088/0067-0049/197/2/36},
  \href {https://ui.adsabs.harvard.edu/abs/2011ApJS..197...36K} {197, 36}

\bibitem[\protect\citeauthoryear{{Kreckel} et~al.,}{{Kreckel}
  et~al.}{2020}]{2020MNRAS.499..193K}
{Kreckel} K.,  et~al., 2020, \mn@doi [\mnras] {10.1093/mnras/staa2743}, \href
  {https://ui.adsabs.harvard.edu/abs/2020MNRAS.499..193K} {499, 193}

\bibitem[\protect\citeauthoryear{{Krumholz} \& {Ting}}{{Krumholz} \&
  {Ting}}{2018}]{2018MNRAS.475.2236K}
{Krumholz} M.~R.,  {Ting} Y.-S.,  2018, \mn@doi [\mnras]
  {10.1093/mnras/stx3286}, \href
  {https://ui.adsabs.harvard.edu/abs/2018MNRAS.475.2236K} {475, 2236}

\bibitem[\protect\citeauthoryear{{Krumholz}, {Burkhart}, {Forbes}  \&
  {Crocker}}{{Krumholz} et~al.}{2018}]{2018MNRAS.477.2716K}
{Krumholz} M.~R.,  {Burkhart} B.,  {Forbes} J.~C.,   {Crocker} R.~M.,  2018,
  \mn@doi [\mnras] {10.1093/mnras/sty852}, \href
  {https://ui.adsabs.harvard.edu/abs/2018MNRAS.477.2716K} {477, 2716}

\bibitem[\protect\citeauthoryear{{Lang} et~al.,}{{Lang}
  et~al.}{2017}]{2017ApJ...840...92L}
{Lang} P.,  et~al., 2017, \mn@doi [\apj] {10.3847/1538-4357/aa6d82}, \href
  {https://ui.adsabs.harvard.edu/abs/2017ApJ...840...92L} {840, 92}

\bibitem[\protect\citeauthoryear{{Larkin} et~al.,}{{Larkin}
  et~al.}{2006}]{2006SPIE.6269E..1AL}
{Larkin} J.,  et~al., 2006, in \procspie. p. 62691A, \mn@doi{10.1117/12.672061}

\bibitem[\protect\citeauthoryear{{Leethochawalit}, {Jones}, {Ellis}, {Stark},
  {Richard}, {Zitrin}  \& {Auger}}{{Leethochawalit}
  et~al.}{2016}]{2016ApJ...820...84L}
{Leethochawalit} N.,  {Jones} T.~A.,  {Ellis} R.~S.,  {Stark} D.~P.,  {Richard}
  J.,  {Zitrin} A.,   {Auger} M.,  2016, \mn@doi [\apj]
  {10.3847/0004-637X/820/2/84}, \href
  {https://ui.adsabs.harvard.edu/abs/2016ApJ...820...84L} {820, 84}

\bibitem[\protect\citeauthoryear{{Li}, {Krumholz}, {Wisnioski}, {Mendel},
  {Kewley}, {S{\'a}nchez}  \& {Galbany}}{{Li}
  et~al.}{2021}]{2021MNRAS.504.5496L}
{Li} Z.,  {Krumholz} M.~R.,  {Wisnioski} E.,  {Mendel} J.~T.,  {Kewley} L.~J.,
  {S{\'a}nchez} S.~F.,   {Galbany} L.,  2021, \mn@doi [\mnras]
  {10.1093/mnras/stab1263}, \href
  {https://ui.adsabs.harvard.edu/abs/2021MNRAS.504.5496L} {504, 5496}

\bibitem[\protect\citeauthoryear{{Ma}, {Hopkins}, {Feldmann}, {Torrey},
  {Faucher-Gigu{\`e}re}  \& {Kere{\v{s}}}}{{Ma}
  et~al.}{2017}]{2017MNRAS.466.4780M}
{Ma} X.,  {Hopkins} P.~F.,  {Feldmann} R.,  {Torrey} P.,  {Faucher-Gigu{\`e}re}
  C.-A.,   {Kere{\v{s}}} D.,  2017, \mn@doi [\mnras] {10.1093/mnras/stx034},
  \href {https://ui.adsabs.harvard.edu/abs/2017MNRAS.466.4780M} {466, 4780}

\bibitem[\protect\citeauthoryear{{Maiolino} \& {Mannucci}}{{Maiolino} \&
  {Mannucci}}{2019}]{2019A&ARv..27....3M}
{Maiolino} R.,  {Mannucci} F.,  2019, \mn@doi [\aapr]
  {10.1007/s00159-018-0112-2}, \href
  {https://ui.adsabs.harvard.edu/abs/2019A&ARv..27....3M} {27, 3}

\bibitem[\protect\citeauthoryear{{Mast} et~al.,}{{Mast}
  et~al.}{2014}]{2014A&A...561A.129M}
{Mast} D.,  et~al., 2014, \mn@doi [\aap] {10.1051/0004-6361/201321789}, \href
  {https://ui.adsabs.harvard.edu/abs/2014A&A...561A.129M} {561, A129}

\bibitem[\protect\citeauthoryear{{McDermid} et~al.,}{{McDermid}
  et~al.}{2020}]{2020arXiv200909242M}
{McDermid} R.~M.,  et~al., 2020, arXiv e-prints, \href
  {https://ui.adsabs.harvard.edu/abs/2020arXiv200909242M} {p. arXiv:2009.09242}

\bibitem[\protect\citeauthoryear{{McGaugh}, {Schombert}, {Bothun}  \& {de
  Blok}}{{McGaugh} et~al.}{2000}]{2000ApJ...533L..99M}
{McGaugh} S.~S.,  {Schombert} J.~M.,  {Bothun} G.~D.,   {de Blok} W.~J.~G.,
  2000, \mn@doi [\apjl] {10.1086/312628}, \href
  {https://ui.adsabs.harvard.edu/abs/2000ApJ...533L..99M} {533, L99}

\bibitem[\protect\citeauthoryear{{McGregor} et~al.,}{{McGregor}
  et~al.}{2003}]{2003SPIE.4841.1581M}
{McGregor} P.~J.,  et~al., 2003, in {Iye} M.,  {Moorwood} A. F.~M.,  eds,
  Society of Photo-Optical Instrumentation Engineers (SPIE) Conference Series
  Vol. 4841, Instrument Design and Performance for Optical/Infrared
  Ground-based Telescopes. pp 1581--1591, \mn@doi{10.1117/12.459448}

\bibitem[\protect\citeauthoryear{{Moster}, {Somerville}, {Maulbetsch}, {van den
  Bosch}, {Macci{\`o}}, {Naab}  \& {Oser}}{{Moster}
  et~al.}{2010}]{2010ApJ...710..903M}
{Moster} B.~P.,  {Somerville} R.~S.,  {Maulbetsch} C.,  {van den Bosch} F.~C.,
  {Macci{\`o}} A.~V.,  {Naab} T.,   {Oser} L.,  2010, \mn@doi [\apj]
  {10.1088/0004-637X/710/2/903}, \href
  {https://ui.adsabs.harvard.edu/abs/2010ApJ...710..903M} {710, 903}

\bibitem[\protect\citeauthoryear{{Mowla}, {van der Wel}, {van Dokkum}  \&
  {Miller}}{{Mowla} et~al.}{2019}]{2019ApJ...872L..13M}
{Mowla} L.,  {van der Wel} A.,  {van Dokkum} P.,   {Miller} T.~B.,  2019,
  \mn@doi [\apjl] {10.3847/2041-8213/ab0379}, \href
  {https://ui.adsabs.harvard.edu/abs/2019ApJ...872L..13M} {872, L13}

\bibitem[\protect\citeauthoryear{{Muratov}, {Kere{\v{s}}},
  {Faucher-Gigu{\`e}re}, {Hopkins}, {Quataert}  \& {Murray}}{{Muratov}
  et~al.}{2015}]{2015MNRAS.454.2691M}
{Muratov} A.~L.,  {Kere{\v{s}}} D.,  {Faucher-Gigu{\`e}re} C.-A.,  {Hopkins}
  P.~F.,  {Quataert} E.,   {Murray} N.,  2015, \mn@doi [\mnras]
  {10.1093/mnras/stv2126}, \href
  {https://ui.adsabs.harvard.edu/abs/2015MNRAS.454.2691M} {454, 2691}

\bibitem[\protect\citeauthoryear{{Nelson} et~al.,}{{Nelson}
  et~al.}{2016}]{2016ApJ...828...27N}
{Nelson} E.~J.,  et~al., 2016, \mn@doi [\apj] {10.3847/0004-637X/828/1/27},
  \href {https://ui.adsabs.harvard.edu/abs/2016ApJ...828...27N} {828, 27}

\bibitem[\protect\citeauthoryear{{Newman} et~al.,}{{Newman}
  et~al.}{2014}]{2014ApJ...781...21N}
{Newman} S.~F.,  et~al., 2014, \mn@doi [\apj] {10.1088/0004-637X/781/1/21},
  \href {https://ui.adsabs.harvard.edu/abs/2014ApJ...781...21N} {781, 21}

\bibitem[\protect\citeauthoryear{{Oliphant}}{{Oliphant}}{2006}]{oliphant2006guide}
{Oliphant} T.~E.,  2006, A guide to NumPy.
~ Vol. 1, Trelgol Publishing USA

\bibitem[\protect\citeauthoryear{{Ostriker} \& {Shetty}}{{Ostriker} \&
  {Shetty}}{2011}]{2011ApJ...731...41O}
{Ostriker} E.~C.,  {Shetty} R.,  2011, \mn@doi [\apj]
  {10.1088/0004-637X/731/1/41}, \href
  {https://ui.adsabs.harvard.edu/abs/2011ApJ...731...41O} {731, 41}

\bibitem[\protect\citeauthoryear{{Pasquini} et~al.,}{{Pasquini}
  et~al.}{2002}]{2002Msngr.110....1P}
{Pasquini} L.,  et~al., 2002, The Messenger, \href
  {https://ui.adsabs.harvard.edu/abs/2002Msngr.110....1P} {110, 1}

\bibitem[\protect\citeauthoryear{Patankar}{Patankar}{1980}]{patankar1980numerical}
Patankar S.~V.,  1980, Numerical heat transfer and fluid flow.
Series on Computational Methods in Mechanics and Thermal Science, Hemisphere
  Publishing Corporation (CRC Press, Taylor \& Francis Group), \url
  {http://www.crcpress.com/product/isbn/9780891165224}

\bibitem[\protect\citeauthoryear{{P{\'e}rez-Montero} \&
  {Contini}}{{P{\'e}rez-Montero} \& {Contini}}{2009}]{2009MNRAS.398..949P}
{P{\'e}rez-Montero} E.,  {Contini} T.,  2009, \mn@doi [\mnras]
  {10.1111/j.1365-2966.2009.15145.x}, \href
  {https://ui.adsabs.harvard.edu/abs/2009MNRAS.398..949P} {398, 949}

\bibitem[\protect\citeauthoryear{{Petit}, {Krumholz}, {Goldbaum}  \&
  {Forbes}}{{Petit} et~al.}{2015}]{2015MNRAS.449.2588P}
{Petit} A.~C.,  {Krumholz} M.~R.,  {Goldbaum} N.~J.,   {Forbes} J.~C.,  2015,
  \mn@doi [\mnras] {10.1093/mnras/stv493}, \href
  {https://ui.adsabs.harvard.edu/abs/2015MNRAS.449.2588P} {449, 2588}

\bibitem[\protect\citeauthoryear{{Pettini} \& {Pagel}}{{Pettini} \&
  {Pagel}}{2004}]{2004MNRAS.348L..59P}
{Pettini} M.,  {Pagel} B. E.~J.,  2004, \mn@doi [\mnras]
  {10.1111/j.1365-2966.2004.07591.x}, \href
  {https://ui.adsabs.harvard.edu/abs/2004MNRAS.348L..59P} {348, L59}

\bibitem[\protect\citeauthoryear{{Pillepich} et~al.,}{{Pillepich}
  et~al.}{2018}]{2018MNRAS.473.4077P}
{Pillepich} A.,  et~al., 2018, \mn@doi [\mnras] {10.1093/mnras/stx2656}, \href
  {https://ui.adsabs.harvard.edu/abs/2018MNRAS.473.4077P} {473, 4077}

\bibitem[\protect\citeauthoryear{{Poetrodjojo} et~al.,}{{Poetrodjojo}
  et~al.}{2021}]{10.1093/mnras/stab205}
{Poetrodjojo} H.,  et~al., 2021, \mn@doi [\mnras] {10.1093/mnras/stab205},
  \href {https://ui.adsabs.harvard.edu/abs/2021MNRAS.502.3357P} {502, 3357}

\bibitem[\protect\citeauthoryear{{Queyrel} et~al.,}{{Queyrel}
  et~al.}{2012}]{2012A&A...539A..93Q}
{Queyrel} J.,  et~al., 2012, \mn@doi [\aap] {10.1051/0004-6361/201117718},
  \href {https://ui.adsabs.harvard.edu/abs/2012A&A...539A..93Q} {539, A93}

\bibitem[\protect\citeauthoryear{{Rich}, {Torrey}, {Kewley}, {Dopita}  \&
  {Rupke}}{{Rich} et~al.}{2012}]{2012ApJ...753....5R}
{Rich} J.~A.,  {Torrey} P.,  {Kewley} L.~J.,  {Dopita} M.~A.,   {Rupke}
  D.~S.~N.,  2012, \mn@doi [\apj] {10.1088/0004-637X/753/1/5}, \href
  {https://ui.adsabs.harvard.edu/abs/2012ApJ...753....5R} {753, 5}

\bibitem[\protect\citeauthoryear{{Richardson}, {Routledge}, {Thatte}, {Tecza},
  {Houghton}, {Pereira-Santaella}  \& {Rigopoulou}}{{Richardson}
  et~al.}{2020}]{2020MNRAS.498.1891R}
{Richardson} M. L.~A.,  {Routledge} L.,  {Thatte} N.,  {Tecza} M.,  {Houghton}
  R. C.~W.,  {Pereira-Santaella} M.,   {Rigopoulou} D.,  2020, \mn@doi [\mnras]
  {10.1093/mnras/staa2317}, \href
  {https://ui.adsabs.harvard.edu/abs/2020MNRAS.498.1891R} {498, 1891}

\bibitem[\protect\citeauthoryear{{Rupke}, {Kewley}  \& {Chien}}{{Rupke}
  et~al.}{2010}]{2010ApJ...723.1255R}
{Rupke} D. S.~N.,  {Kewley} L.~J.,   {Chien} L.~H.,  2010, \mn@doi [\apj]
  {10.1088/0004-637X/723/2/1255}, \href
  {https://ui.adsabs.harvard.edu/abs/2010ApJ...723.1255R} {723, 1255}

\bibitem[\protect\citeauthoryear{{S{\'a}nchez}}{{S{\'a}nchez}}{2020}]{2020ARA&A..58...99S}
{S{\'a}nchez} S.~F.,  2020, \mn@doi [\araa]
  {10.1146/annurev-astro-012120-013326}, \href
  {https://ui.adsabs.harvard.edu/abs/2020ARA&A..58...99S} {58, 99}

\bibitem[\protect\citeauthoryear{{S{\'a}nchez}, {Walcher}, {Lopez-Cob{\'a}},
  {Barrera-Ballesteros}, {Mej{\'\i}a-Narv{\'a}ez}, {Espinosa-Ponce}  \&
  {Camps-Fari{\~n}a}}{{S{\'a}nchez} et~al.}{2021}]{2020arXiv200900424S}
{S{\'a}nchez} S.~F.,  {Walcher} C.~J.,  {Lopez-Cob{\'a}} C.,
  {Barrera-Ballesteros} J.~K.,  {Mej{\'\i}a-Narv{\'a}ez} A.,  {Espinosa-Ponce}
  C.,   {Camps-Fari{\~n}a} A.,  2021, \mn@doi [\rmxaa]
  {10.22201/ia.01851101p.2021.57.01.01}, \href
  {https://ui.adsabs.harvard.edu/abs/2021RMxAA..57....3S} {57, 3}

\bibitem[\protect\citeauthoryear{{Searle}}{{Searle}}{1971}]{1971ApJ...168..327S}
{Searle} L.,  1971, \mn@doi [\apj] {10.1086/151090}, \href
  {https://ui.adsabs.harvard.edu/abs/1971ApJ...168..327S} {168, 327}

\bibitem[\protect\citeauthoryear{{Sharda}, {Krumholz}, {Wisnioski}, {Forbes},
  {Federrath}  \& {Acharyya}}{{Sharda} et~al.}{2021a}]{2020aMNRAS.xxx..xxxS}
{Sharda} P.,  {Krumholz} M.~R.,  {Wisnioski} E.,  {Forbes} J.~C.,  {Federrath}
  C.,   {Acharyya} A.,  2021a, \mn@doi [\mnras] {10.1093/mnras/stab252}, \href
  {https://ui.adsabs.harvard.edu/abs/2021MNRAS.502.5935S} {502, 5935}

\bibitem[\protect\citeauthoryear{{Sharda}, {Krumholz}, {Wisnioski}, {Acharyya},
  {Federrath}  \& {Forbes}}{{Sharda} et~al.}{2021b}]{2020bMNRAS.xxx..xxxS}
{Sharda} P.,  {Krumholz} M.~R.,  {Wisnioski} E.,  {Acharyya} A.,  {Federrath}
  C.,   {Forbes} J.~C.,  2021b, \mn@doi [\mnras] {10.1093/mnras/stab868}, \href
  {https://ui.adsabs.harvard.edu/abs/2021MNRAS.504...53S} {504, 53}

\bibitem[\protect\citeauthoryear{{Sharples} et~al.,}{{Sharples}
  et~al.}{2004}]{2004SPIE.5492.1179S}
{Sharples} R.~M.,  et~al., 2004, in {Moorwood} A. F.~M.,  {Iye} M.,  eds,
  Society of Photo-Optical Instrumentation Engineers (SPIE) Conference Series
  Vol. 5492, \procspie. pp 1179--1186, \mn@doi{10.1117/12.550495}

\bibitem[\protect\citeauthoryear{{Shaver}, {McGee}, {Newton}, {Danks}  \&
  {Pottasch}}{{Shaver} et~al.}{1983}]{1983MNRAS.204...53S}
{Shaver} P.~A.,  {McGee} R.~X.,  {Newton} L.~M.,  {Danks} A.~C.,   {Pottasch}
  S.~R.,  1983, \mn@doi [\mnras] {10.1093/mnras/204.1.53}, \href
  {https://ui.adsabs.harvard.edu/abs/1983MNRAS.204...53S} {204, 53}

\bibitem[\protect\citeauthoryear{{Shirazi}, {Brinchmann}  \&
  {Rahmati}}{{Shirazi} et~al.}{2014}]{2014ApJ...787..120S}
{Shirazi} M.,  {Brinchmann} J.,   {Rahmati} A.,  2014, \mn@doi [\apj]
  {10.1088/0004-637X/787/2/120}, \href
  {https://ui.adsabs.harvard.edu/abs/2014ApJ...787..120S} {787, 120}

\bibitem[\protect\citeauthoryear{{Sillero}, {Tissera}, {Lambas}  \&
  {Michel-Dansac}}{{Sillero} et~al.}{2017}]{2017MNRAS.472.4404S}
{Sillero} E.,  {Tissera} P.~B.,  {Lambas} D.~G.,   {Michel-Dansac} L.,  2017,
  \mn@doi [\mnras] {10.1093/mnras/stx2265}, \href
  {https://ui.adsabs.harvard.edu/abs/2017MNRAS.472.4404S} {472, 4404}

\bibitem[\protect\citeauthoryear{{Simons} et~al.,}{{Simons}
  et~al.}{2017}]{2017ApJ...843...46S}
{Simons} R.~C.,  et~al., 2017, \mn@doi [\apj] {10.3847/1538-4357/aa740c}, \href
  {https://ui.adsabs.harvard.edu/abs/2017ApJ...843...46S} {843, 46}

\bibitem[\protect\citeauthoryear{{Simons} et~al.,}{{Simons}
  et~al.}{2019}]{2019ApJ...874...59S}
{Simons} R.~C.,  et~al., 2019, \mn@doi [\apj] {10.3847/1538-4357/ab07c9}, \href
  {https://ui.adsabs.harvard.edu/abs/2019ApJ...874...59S} {874, 59}

\bibitem[\protect\citeauthoryear{{Simons} et~al.,}{{Simons}
  et~al.}{2020}]{2020arXiv201103553S}
{Simons} R.~C.,  et~al., 2020, arXiv e-prints, \href
  {https://ui.adsabs.harvard.edu/abs/2020arXiv201103553S} {p. arXiv:2011.03553}

\bibitem[\protect\citeauthoryear{{Skelton} et~al.,}{{Skelton}
  et~al.}{2014}]{2014ApJS..214...24S}
{Skelton} R.~E.,  et~al., 2014, \mn@doi [\apjs] {10.1088/0067-0049/214/2/24},
  \href {https://ui.adsabs.harvard.edu/abs/2014ApJS..214...24S} {214, 24}

\bibitem[\protect\citeauthoryear{{Sobral} et~al.,}{{Sobral}
  et~al.}{2009}]{2009MNRAS.398...75S}
{Sobral} D.,  et~al., 2009, \mn@doi [\mnras]
  {10.1111/j.1365-2966.2009.15129.x}, \href
  {https://ui.adsabs.harvard.edu/abs/2009MNRAS.398...75S} {398, 75}

\bibitem[\protect\citeauthoryear{{Sobral}, {Smail}, {Best}, {Geach}, {Matsuda},
  {Stott}, {Cirasuolo}  \& {Kurk}}{{Sobral}
  et~al.}{2013a}]{2013MNRAS.428.1128S}
{Sobral} D.,  {Smail} I.,  {Best} P.~N.,  {Geach} J.~E.,  {Matsuda} Y.,
  {Stott} J.~P.,  {Cirasuolo} M.,   {Kurk} J.,  2013a, \mn@doi [\mnras]
  {10.1093/mnras/sts096}, \href
  {https://ui.adsabs.harvard.edu/abs/2013MNRAS.428.1128S} {428, 1128}

\bibitem[\protect\citeauthoryear{{Sobral} et~al.,}{{Sobral}
  et~al.}{2013b}]{2013ApJ...779..139S}
{Sobral} D.,  et~al., 2013b, \mn@doi [\apj] {10.1088/0004-637X/779/2/139},
  \href {https://ui.adsabs.harvard.edu/abs/2013ApJ...779..139S} {779, 139}

\bibitem[\protect\citeauthoryear{{Springel} \& {Hernquist}}{{Springel} \&
  {Hernquist}}{2003}]{2003MNRAS.339..289S}
{Springel} V.,  {Hernquist} L.,  2003, \mn@doi [\mnras]
  {10.1046/j.1365-8711.2003.06206.x}, \href
  {https://ui.adsabs.harvard.edu/abs/2003MNRAS.339..289S} {339, 289}

\bibitem[\protect\citeauthoryear{{Stott} et~al.,}{{Stott}
  et~al.}{2014}]{2014MNRAS.443.2695S}
{Stott} J.~P.,  et~al., 2014, \mn@doi [\mnras] {10.1093/mnras/stu1343}, \href
  {https://ui.adsabs.harvard.edu/abs/2014MNRAS.443.2695S} {443, 2695}

\bibitem[\protect\citeauthoryear{{Stott} et~al.,}{{Stott}
  et~al.}{2016}]{2016MNRAS.457.1888S}
{Stott} J.~P.,  et~al., 2016, \mn@doi [\mnras] {10.1093/mnras/stw129}, \href
  {https://ui.adsabs.harvard.edu/abs/2016MNRAS.457.1888S} {457, 1888}

\bibitem[\protect\citeauthoryear{{Straatman} et~al.,}{{Straatman}
  et~al.}{2017}]{2017ApJ...839...57S}
{Straatman} C. M.~S.,  et~al., 2017, \mn@doi [\apj] {10.3847/1538-4357/aa643e},
  \href {https://ui.adsabs.harvard.edu/abs/2017ApJ...839...57S} {839, 57}

\bibitem[\protect\citeauthoryear{{Strom}, {Steidel}, {Rudie}, {Trainor}  \&
  {Pettini}}{{Strom} et~al.}{2018}]{2018ApJ...868..117S}
{Strom} A.~L.,  {Steidel} C.~C.,  {Rudie} G.~C.,  {Trainor} R.~F.,   {Pettini}
  M.,  2018, \mn@doi [\apj] {10.3847/1538-4357/aae1a5}, \href
  {https://ui.adsabs.harvard.edu/abs/2018ApJ...868..117S} {868, 117}

\bibitem[\protect\citeauthoryear{{Swinbank}, {Sobral}, {Smail}, {Geach},
  {Best}, {McCarthy}, {Crain}  \& {Theuns}}{{Swinbank}
  et~al.}{2012}]{2012MNRAS.426..935S}
{Swinbank} A.~M.,  {Sobral} D.,  {Smail} I.,  {Geach} J.~E.,  {Best} P.~N.,
  {McCarthy} I.~G.,  {Crain} R.~A.,   {Theuns} T.,  2012, \mn@doi [\mnras]
  {10.1111/j.1365-2966.2012.21774.x}, \href
  {https://ui.adsabs.harvard.edu/abs/2012MNRAS.426..935S} {426, 935}

\bibitem[\protect\citeauthoryear{{Tacconi}, {Genzel}  \& {Sternberg}}{{Tacconi}
  et~al.}{2020}]{2020ARA&A..58..157T}
{Tacconi} L.~J.,  {Genzel} R.,   {Sternberg} A.,  2020, \mn@doi [\araa]
  {10.1146/annurev-astro-082812-141034}, \href
  {https://ui.adsabs.harvard.edu/abs/2020ARA&A..58..157T} {58, 157}

\bibitem[\protect\citeauthoryear{{Teklu}, {Remus}, {Dolag}, {Arth}, {Burkert},
  {Obreja}  \& {Schulze}}{{Teklu} et~al.}{2018}]{2018ApJ...854L..28T}
{Teklu} A.~F.,  {Remus} R.-S.,  {Dolag} K.,  {Arth} A.,  {Burkert} A.,
  {Obreja} A.,   {Schulze} F.,  2018, \mn@doi [\apjl]
  {10.3847/2041-8213/aaaeb4}, \href
  {https://ui.adsabs.harvard.edu/abs/2018ApJ...854L..28T} {854, L28}

\bibitem[\protect\citeauthoryear{{Thatte} et~al.,}{{Thatte}
  et~al.}{2014}]{2014SPIE.9147E..25T}
{Thatte} N.~A.,  et~al., 2014, in {Ramsay} S.~K.,  {McLean} I.~S.,   {Takami}
  H.,  eds,  Society of Photo-Optical Instrumentation Engineers (SPIE)
  Conference Series Vol. 9147, Ground-based and Airborne Instrumentation for
  Astronomy V. p. 914725, \mn@doi{10.1117/12.2055436}

\bibitem[\protect\citeauthoryear{{Tiley} et~al.,}{{Tiley}
  et~al.}{2016}]{2016MNRAS.460..103T}
{Tiley} A.~L.,  et~al., 2016, \mn@doi [\mnras] {10.1093/mnras/stw936}, \href
  {https://ui.adsabs.harvard.edu/abs/2016MNRAS.460..103T} {460, 103}

\bibitem[\protect\citeauthoryear{{Tiley} et~al.,}{{Tiley}
  et~al.}{2019a}]{2019MNRAS.482.2166T}
{Tiley} A.~L.,  et~al., 2019a, \mn@doi [\mnras] {10.1093/mnras/sty2794}, \href
  {https://ui.adsabs.harvard.edu/abs/2019MNRAS.482.2166T} {482, 2166}

\bibitem[\protect\citeauthoryear{{Tiley} et~al.,}{{Tiley}
  et~al.}{2019b}]{2019MNRAS.485..934T}
{Tiley} A.~L.,  et~al., 2019b, \mn@doi [\mnras] {10.1093/mnras/stz428}, \href
  {https://ui.adsabs.harvard.edu/abs/2019MNRAS.485..934T} {485, 934}

\bibitem[\protect\citeauthoryear{{Torres-Flores}, {Scarano}, {Mendes de
  Oliveira}, {de Mello}, {Amram}  \& {Plana}}{{Torres-Flores}
  et~al.}{2014}]{2014MNRAS.438.1894T}
{Torres-Flores} S.,  {Scarano} S.,  {Mendes de Oliveira} C.,  {de Mello} D.~F.,
   {Amram} P.,   {Plana} H.,  2014, \mn@doi [\mnras] {10.1093/mnras/stt2340},
  \href {https://ui.adsabs.harvard.edu/abs/2014MNRAS.438.1894T} {438, 1894}

\bibitem[\protect\citeauthoryear{{Tully} \& {Fisher}}{{Tully} \&
  {Fisher}}{1977}]{1977A&A....54..661T}
{Tully} R.~B.,  {Fisher} J.~R.,  1977, \aap, \href
  {https://ui.adsabs.harvard.edu/abs/1977A&A....54..661T} {500, 105}

\bibitem[\protect\citeauthoryear{{{\"U}bler} et~al.,}{{{\"U}bler}
  et~al.}{2017}]{2017ApJ...842..121U}
{{\"U}bler} H.,  et~al., 2017, \mn@doi [\apj] {10.3847/1538-4357/aa7558}, \href
  {https://ui.adsabs.harvard.edu/abs/2017ApJ...842..121U} {842, 121}

\bibitem[\protect\citeauthoryear{{{\"U}bler} et~al.,}{{{\"U}bler}
  et~al.}{2019}]{2019ApJ...880...48U}
{{\"U}bler} H.,  et~al., 2019, \mn@doi [\apj] {10.3847/1538-4357/ab27cc}, \href
  {https://ui.adsabs.harvard.edu/abs/2019ApJ...880...48U} {880, 48}

\bibitem[\protect\citeauthoryear{{Urrutia} et~al.,}{{Urrutia}
  et~al.}{2019}]{2019A&A...624A.141U}
{Urrutia} T.,  et~al., 2019, \mn@doi [\aap] {10.1051/0004-6361/201834656},
  \href {https://ui.adsabs.harvard.edu/abs/2019A&A...624A.141U} {624, A141}

\bibitem[\protect\citeauthoryear{{Virtanen} et~al.,}{{Virtanen}
  et~al.}{2020}]{2020NatMe..17..261V}
{Virtanen} P.,  et~al., 2020, \mn@doi [Nature Methods]
  {10.1038/s41592-019-0686-2}, \href
  {https://ui.adsabs.harvard.edu/abs/2020NatMe..17..261V} {17, 261}

\bibitem[\protect\citeauthoryear{{Whitaker}, {van Dokkum}, {Brammer}  \&
  {Franx}}{{Whitaker} et~al.}{2012}]{2012ApJ...754L..29W}
{Whitaker} K.~E.,  {van Dokkum} P.~G.,  {Brammer} G.,   {Franx} M.,  2012,
  \mn@doi [\apjl] {10.1088/2041-8205/754/2/L29}, \href
  {https://ui.adsabs.harvard.edu/abs/2012ApJ...754L..29W} {754, L29}

\bibitem[\protect\citeauthoryear{{Wisnioski} et~al.,}{{Wisnioski}
  et~al.}{2015}]{2015ApJ...799..209W}
{Wisnioski} E.,  et~al., 2015, \mn@doi [\apj] {10.1088/0004-637X/799/2/209},
  \href {https://ui.adsabs.harvard.edu/abs/2015ApJ...799..209W} {799, 209}

\bibitem[\protect\citeauthoryear{{Wisnioski} et~al.,}{{Wisnioski}
  et~al.}{2018}]{2018ApJ...855...97W}
{Wisnioski} E.,  et~al., 2018, \mn@doi [\apj] {10.3847/1538-4357/aab097}, \href
  {https://ui.adsabs.harvard.edu/abs/2018ApJ...855...97W} {855, 97}

\bibitem[\protect\citeauthoryear{{Wisnioski} et~al.,}{{Wisnioski}
  et~al.}{2019}]{2019ApJ...886..124W}
{Wisnioski} E.,  et~al., 2019, \mn@doi [\apj] {10.3847/1538-4357/ab4db8}, \href
  {https://ui.adsabs.harvard.edu/abs/2019ApJ...886..124W} {886, 124}

\bibitem[\protect\citeauthoryear{{Wright}}{{Wright}}{2006}]{2006PASP..118.1711W}
{Wright} E.~L.,  2006, \mn@doi [\pasp] {10.1086/510102}, \href
  {https://ui.adsabs.harvard.edu/abs/2006PASP..118.1711W} {118, 1711}

\bibitem[\protect\citeauthoryear{{Wuyts} et~al.,}{{Wuyts}
  et~al.}{2016}]{2016ApJ...827...74W}
{Wuyts} E.,  et~al., 2016, \mn@doi [\apj] {10.3847/0004-637X/827/1/74}, \href
  {https://ui.adsabs.harvard.edu/abs/2016ApJ...827...74W} {827, 74}

\bibitem[\protect\citeauthoryear{{Yang} \& {Krumholz}}{{Yang} \&
  {Krumholz}}{2012}]{2012ApJ...758...48Y}
{Yang} C.-C.,  {Krumholz} M.,  2012, \mn@doi [\apj]
  {10.1088/0004-637X/758/1/48}, \href
  {https://ui.adsabs.harvard.edu/abs/2012ApJ...758...48Y} {758, 48}

\bibitem[\protect\citeauthoryear{{Yuan}, {Kewley}, {Swinbank}, {Richard}  \&
  {Livermore}}{{Yuan} et~al.}{2011}]{2011ApJ...732L..14Y}
{Yuan} T.~T.,  {Kewley} L.~J.,  {Swinbank} A.~M.,  {Richard} J.,   {Livermore}
  R.~C.,  2011, \mn@doi [\apjl] {10.1088/2041-8205/732/1/L14}, \href
  {https://ui.adsabs.harvard.edu/abs/2011ApJ...732L..14Y} {732, L14}

\bibitem[\protect\citeauthoryear{{Yuan}, {Kewley}  \& {Rich}}{{Yuan}
  et~al.}{2013}]{2013ApJ...767..106Y}
{Yuan} T.~T.,  {Kewley} L.~J.,   {Rich} J.,  2013, \mn@doi [\apj]
  {10.1088/0004-637X/767/2/106}, \href
  {https://ui.adsabs.harvard.edu/abs/2013ApJ...767..106Y} {767, 106}

\bibitem[\protect\citeauthoryear{{van der Wel} et~al.,}{{van der Wel}
  et~al.}{2012}]{2012ApJS..203...24V}
{van der Wel} A.,  et~al., 2012, \mn@doi [\apjs] {10.1088/0067-0049/203/2/24},
  \href {https://ui.adsabs.harvard.edu/abs/2012ApJS..203...24V} {203, 24}

\makeatother
\end{thebibliography}


\appendix

\section{Compiled data with reanalysed kinematics}
\label{s:app_data}
\onecolumn
\begin{longtable}{|lcrrcrrrr|} \caption{Compiled database of metallicity gradients and reanalysed kinematics utilized in this work. Columns $1-9$ list the parent sample, galaxy ID, redshift, stellar mass, star formation rate, half-light radius (in arcsec), metallicity gradient, rotational velocity, and velocity dispersion, respectively. See \autoref{s:data} and \autoref{tab:tab1} for the list of references for each sample.} 
\label{tab:tab2}\\
\hline
Sample & Galaxy ID & $z$ & $\log_{10}\,\frac{M_{\star}}{M_{\odot}}$ & $\log_{10}\,\frac{\mathrm{SFR}}{M_{\odot}\,\mathrm{yr^{-1}}}$ & $r_{\mathrm{e}}\,\mathrm{(\arcsec)}$ & $\frac{\nabla \left(\log_{10} \mathcal{Z}\right)}{\mathrm{dex\,kpc^{-1}}}$ & $v_{\phi}/\mathrm{km\,s^{-1}}$ & $\sigma_{\rm{g}}/\mathrm{km\,s^{-1}}$\\
(1) & (2) & (3) & (4) & (5) & (6) & (7) & (8) & (9) \\
\hline
\textit{MASSIV}	&	20106882	&	1.40	&	9.73	&	1.58	&	0.42	&$	0.033	_{	-0.013	}^{+	0.013	}$&$	103	_{-	13	}^{+	13	}$&$	46	_{-	29	}^{+	24	}$\\
\textit{MASSIV}	&	20116027	&	1.53	&	9.83	&	1.63	&	0.50	&$	0.023	_{	-0.012	}^{+	0.012	}$&$	19	_{-	5	}^{+	5	}$&$	37	_{-	21	}^{+	21	}$\\
\textit{MASSIV}	&	20147106	&	1.52	&	9.84	&	1.96	&	0.14	&$	0.021	_{	-0.027	}^{+	0.027	}$&$	9	_{-	6	}^{+	6	}$&$	12	_{-	6	}^{+	11	}$\\
\textit{MASSIV}	&	20214655	&	1.04	&	9.76	&	1.71	&	0.18	&$	0.003	_{	-0.008	}^{+	0.008	}$&$	19	_{-	8	}^{+	8	}$&$	7	_{-	4	}^{+	8	}$\\
\textit{MASSIV}	&	20386743	&	1.05	&	9.62	&	1.60	&	0.33	&$	0.006	_{	-0.012	}^{+	0.012	}$&$	14	_{-	6	}^{+	6	}$&$	22	_{-	12	}^{+	10	}$\\
\textit{MASSIV}	&	20461235	&	1.03	&	10.10	&	0.98	&	0.49	&$	-0.059	_{	-0.012	}^{+	0.012	}$&$	80	_{-	17	}^{+	17	}$&$	22	_{-	14	}^{+	14	}$\\
\textit{MASSIV}	&	140083410	&	0.94	&	9.81	&	1.57	&	0.24	&$	-0.011	_{	-0.021	}^{+	0.021	}$&$	16	_{-	11	}^{+	11	}$&$	29	_{-	11	}^{+	16	}$\\
\textit{MASSIV}	&	140217425	&	0.98	&	10.58	&	2.30	&	1.12	&$	0.027	_{	-0.002	}^{+	0.002	}$&$	203	_{-	23	}^{+	23	}$&$	66	_{-	35	}^{+	35	}$\\
\textit{MASSIV}	&	140545062	&	1.04	&	10.34	&	1.36	&	0.36	&$	-0.027	_{	-0.006	}^{+	0.006	}$&$	100	_{-	23	}^{+	23	}$&$	33	_{-	14	}^{+	16	}$\\
\textit{MASSIV}	&	220014252	&	1.31	&	10.52	&	2.30	&	0.40	&$	0.016	_{	-0.004	}^{+	0.004	}$&$	59	_{-	14	}^{+	14	}$&$	53	_{-	28	}^{+	28	}$\\
\textit{MASSIV}	&	220015726	&	1.29	&	10.51	&	2.03	&	0.33	&$	0.000	_{	-0.010	}^{+	0.010	}$&$	128	_{-	48	}^{+	48	}$&$	53	_{-	22	}^{+	22	}$\\
\textit{MASSIV}	&	220376206	&	1.24	&	10.41	&	2.40	&	0.63	&$	0.024	_{	-0.011	}^{+	0.011	}$&$	114	_{-	14	}^{+	14	}$&$	66	_{-	31	}^{+	32	}$\\
\textit{MASSIV}	&	220397579	&	1.04	&	9.97	&	2.16	&	0.37	&$	-0.010	_{	-0.010	}^{+	0.010	}$&$	21	_{-	5	}^{+	5	}$&$	37	_{-	18	}^{+	16	}$\\
\textit{MASSIV}	&	220544103	&	1.40	&	10.45	&	2.07	&	0.67	&$	0.005	_{	-0.006	}^{+	0.006	}$&$	97	_{-	28	}^{+	28	}$&$	33	_{-	18	}^{+	19	}$\\
\textit{MASSIV}	&	220544394	&	1.01	&	10.08	&	1.70	&	0.42	&$	0.023	_{	-0.011	}^{+	0.011	}$&$	78	_{-	19	}^{+	19	}$&$	28	_{-	16	}^{+	17	}$\\
\textit{MASSIV}	&	220576226	&	1.02	&	10.05	&	1.82	&	0.27	&$	-0.006	_{	-0.005	}^{+	0.005	}$&$	18	_{-	6	}^{+	6	}$&$	46	_{-	21	}^{+	22	}$\\
\textit{MASSIV}	&	220578040	&	1.05	&	10.46	&	1.32	&	0.47	&$	0.002	_{	-0.007	}^{+	0.007	}$&$	245	_{-	103	}^{+	103	}$&$	56	_{-	26	}^{+	28	}$\\
\textit{MASSIV}	&	220584167	&	1.47	&	10.95	&	2.31	&	0.85	&$	-0.069	_{	-0.007	}^{+	0.007	}$&$	207	_{-	42	}^{+	42	}$&$	42	_{-	18	}^{+	18	}$\\
\textit{MASSIV}	&	910193711	&	1.56	&	9.73	&	2.30	&	0.27	&$	-0.019	_{	-0.014	}^{+	0.014	}$&$	76	_{-	16	}^{+	16	}$&$	91	_{-	27	}^{+	37	}$\\
\textit{HiZELS}	&	CFHT-NBJ-1709	&	0.81	&	10.70	&	0.93	&	0.27	&$	0.007	_{	-0.006	}^{+	0.006	}$&$	125	_{-	33	}^{+	33	}$&$	10	_{-	4	}^{+	6	}$\\
\textit{HiZELS}	&	CFHT-NBJ-1739	&	0.80	&	10.60	&	1.06	&	0.79	&$	-0.001	_{	-0.006	}^{+	0.006	}$&$	257	_{-	51	}^{+	51	}$&$	43	_{-	23	}^{+	23	}$\\
\textit{HiZELS}	&	CFHT-NBJ-1740	&	0.81	&	10.40	&	0.95	&	0.65	&$	0.016	_{	-0.005	}^{+	0.005	}$&$	285	_{-	60	}^{+	60	}$&$	49	_{-	23	}^{+	23	}$\\
\textit{HiZELS}	&	CFHT-NBJ-1745	&	0.82	&	9.80	&	0.75	&	0.54	&$	0.025	_{	-0.009	}^{+	0.009	}$&$	264	_{-	48	}^{+	47	}$&$	58	_{-	26	}^{+	26	}$\\
\textit{HiZELS}	&	CFHT-NBJ-1759	&	0.80	&	10.30	&	1.11	&	0.54	&$	-0.018	_{	-0.003	}^{+	0.003	}$&$	302	_{-	44	}^{+	44	}$&$	15	_{-	7	}^{+	9	}$\\
\textit{HiZELS}	&	CFHT-NBJ-1774	&	0.81	&	9.80	&	0.62	&	0.50	&$	0.013	_{	-0.006	}^{+	0.006	}$&$	82	_{-	24	}^{+	24	}$&$	86	_{-	32	}^{+	37	}$\\
\textit{HiZELS}	&	CFHT-NBJ-1787	&	0.81	&	10.60	&	1.08	&	0.85	&$	0.007	_{	-0.004	}^{+	0.004	}$&$	303	_{-	27	}^{+	27	}$&$	22	_{-	10	}^{+	10	}$\\
\textit{HiZELS}	&	CFHT-NBJ-1790	&	0.81	&	9.90	&	0.67	&	0.22	&$	0.032	_{	-0.006	}^{+	0.006	}$&$	111	_{-	31	}^{+	31	}$&$	10	_{-	4	}^{+	7	}$\\
\textit{HiZELS}	&	CFHT-NBJ-1795	&	0.81	&	9.80	&	0.81	&	0.39	&$	-0.063	_{	-0.010	}^{+	0.010	}$&$	96	_{-	26	}^{+	26	}$&$	10	_{-	4	}^{+	6	}$\\
\textit{SHiZELS}	&	HiZELS1	&	0.84	&	10.03	&	0.30	&	0.23	&$	-0.037	_{	-0.058	}^{+	0.030	}$&$	100	_{-	26	}^{+	26	}$&$	57	_{-	23	}^{+	23	}$\\
\textit{SHiZELS}	&	HiZELS7	&	1.46	&	9.81	&	0.90	&	0.43	&$	-0.019	_{	-0.040	}^{+	0.019	}$&$	106	_{-	18	}^{+	18	}$&$	4	_{-	1	}^{+	2	}$\\
\textit{SHiZELS}	&	HiZELS8	&	1.46	&	10.32	&	0.85	&	0.36	&$	0.006	_{	-0.004	}^{+	0.017	}$&$	190	_{-	26	}^{+	26	}$&$	34	_{-	18	}^{+	18	}$\\
\textit{SHiZELS}	&	HiZELS9	&	1.46	&	10.08	&	0.78	&	0.48	&$	-0.027	_{	-0.018	}^{+	0.010	}$&$	151	_{-	46	}^{+	46	}$&$	59	_{-	29	}^{+	29	}$\\
\textit{SHiZELS}	&	HiZELS10	&	1.45	&	9.42	&	1.00	&	0.27	&$	-0.031	_{	-0.014	}^{+	0.016	}$&$	37	_{-	19	}^{+	19	}$&$	53	_{-	25	}^{+	25	}$\\
\textit{SHiZELS}	&	HiZELS11	&	1.49	&	11.01	&	0.90	&	0.15	&$	-0.087	_{	-0.006	}^{+	0.032	}$&$	311	_{-	42	}^{+	42	}$&$	109	_{-	33	}^{+	39	}$\\
\textit{MUSE-WIDE}	&	G103012059	&	0.56	&	10.25	&	1.37	&	0.72	&$	-0.075	_{	-0.017	}^{+	0.014	}$&$	223	_{-	40	}^{+	40	}$&$	43	_{-	18	}^{+	18	}$\\
\textit{MUSE-WIDE}	&	G118011046	&	0.58	&	10.54	&	1.28	&	1.36	&$	-0.038	_{	-0.003	}^{+	0.003	}$&$	275	_{-	29	}^{+	29	}$&$	37	_{-	15	}^{+	15	}$\\
\textit{MUSE-WIDE}	&	G105002016	&	0.34	&	10.31	&	0.25	&	0.45	&$	-0.025	_{	-0.014	}^{+	0.019	}$&$	141	_{-	11	}^{+	11	}$&$	64	_{-	16	}^{+	14	}$\\
\textit{MUSE-WIDE}	&	G122003050	&	0.21	&	9.97	&	0.02	&	1.44	&$	-0.065	_{	-0.001	}^{+	0.001	}$&$	169	_{-	37	}^{+	37	}$&$	43	_{-	6	}^{+	6	}$\\
\textit{MUSE-WIDE}	&	G102021103	&	0.25	&	9.22	&	-0.68	&	0.78	&$	-0.024	_{	-0.021	}^{+	0.024	}$&$	62	_{-	15	}^{+	15	}$&$	41	_{-	6	}^{+	6	}$\\
\textit{MUSE-WIDE}	&	G105012048	&	0.68	&	10.39	&	1.07	&	0.92	&$	-0.044	_{	-0.008	}^{+	0.006	}$&$	149	_{-	27	}^{+	27	}$&$	46	_{-	19	}^{+	19	}$\\
\textit{MUSE-WIDE}	&	G104005033	&	0.40	&	8.64	&	-0.72	&	0.90	&$	0.047	_{	-0.021	}^{+	0.021	}$&$	43	_{-	10	}^{+	10	}$&$	33	_{-	10	}^{+	10	}$\\
\textit{MUSE-WIDE}	&	G107029135	&	0.74	&	9.75	&	0.84	&	0.50	&$	-0.144	_{	-0.024	}^{+	0.024	}$&$	39	_{-	22	}^{+	22	}$&$	35	_{-	14	}^{+	14	}$\\
\textit{MUSE-WIDE}	&	G101001006	&	0.31	&	8.64	&	-0.65	&	0.64	&$	-0.049	_{	-0.011	}^{+	0.010	}$&$	61	_{-	28	}^{+	28	}$&$	31	_{-	5	}^{+	5	}$\\
\textit{MUSE-WIDE}	&	G114007070	&	0.13	&	8.66	&	-1.50	&	1.46	&$	-0.131	_{	-0.069	}^{+	0.080	}$&$	91	_{-	22	}^{+	22	}$&$	38	_{-	6	}^{+	6	}$\\
\textit{MUSE-WIDE}	&	G108016127	&	0.21	&	8.50	&	-1.05	&	0.62	&$	-0.188	_{	-0.061	}^{+	0.046	}$&$	62	_{-	19	}^{+	19	}$&$	46	_{-	10	}^{+	13	}$\\
\textit{MUSE-WIDE}	&	HDFS3	&	0.56	&	9.75	&	1.39	&	1.34	&$	-0.032	_{	-0.002	}^{+	0.002	}$&$	68	_{-	17	}^{+	17	}$&$	24	_{-	13	}^{+	13	}$\\
\textit{MUSE-WIDE}	&	HDFS6	&	0.42	&	9.40	&	-0.02	&	0.60	&$	0.002	_{	-0.006	}^{+	0.005	}$&$	37	_{-	12	}^{+	12	}$&$	33	_{-	10	}^{+	10	}$\\
\textit{MUSE-WIDE}	&	HDFS7	&	0.46	&	9.49	&	0.19	&	0.73	&$	-0.133	_{	-0.007	}^{+	0.007	}$&$	50	_{-	17	}^{+	17	}$&$	40	_{-	12	}^{+	12	}$\\
\textit{MUSE-WIDE}	&	HDFS8	&	0.58	&	10.00	&	1.12	&	0.30	&$	-0.058	_{	-0.014	}^{+	0.013	}$&$	69	_{-	21	}^{+	21	}$&$	53	_{-	27	}^{+	27	}$\\
\textit{MUSE-WIDE}	&	HDFS9	&	0.56	&	9.49	&	0.96	&	0.42	&$	-0.124	_{	-0.010	}^{+	0.009	}$&$	72	_{-	33	}^{+	33	}$&$	45	_{-	22	}^{+	22	}$\\
\textit{MUSE-WIDE}	&	HDFS11	&	0.58	&	9.31	&	0.52	&	0.17	&$	-0.262	_{	-0.028	}^{+	0.052	}$&$	8	_{-	4	}^{+	15	}$&$	21	_{-	9	}^{+	9	}$\\
\textit{MUSE-WIDE}	&	UDF1	&	0.62	&	10.60	&	1.19	&	1.34	&$	-0.026	_{	-0.003	}^{+	0.003	}$&$	131	_{-	13	}^{+	13	}$&$	34	_{-	15	}^{+	15	}$\\
\textit{MUSE-WIDE}	&	UDF2	&	0.42	&	9.89	&	-0.07	&	0.58	&$	0.002	_{	-0.003	}^{+	0.003	}$&$	119	_{-	12	}^{+	12	}$&$	50	_{-	4	}^{+	4	}$\\
\textit{MUSE-WIDE}	&	UDF3	&	0.62	&	10.13	&	0.82	&	0.86	&$	-0.081	_{	-0.013	}^{+	0.012	}$&$	75	_{-	25	}^{+	25	}$&$	45	_{-	22	}^{+	22	}$\\
\textit{MUSE-WIDE}	&	UDF4	&	0.77	&	10.06	&	1.26	&	0.98	&$	-0.037	_{	-0.003	}^{+	0.003	}$&$	47	_{-	19	}^{+	19	}$&$	37	_{-	16	}^{+	16	}$\\
\textit{MUSE-WIDE}	&	UDF7	&	0.62	&	9.39	&	0.70	&	0.68	&$	-0.188	_{	-0.010	}^{+	0.011	}$&$	10	_{-	3	}^{+	5	}$&$	27	_{-	13	}^{+	13	}$\\
\textit{MUSE-WIDE}	&	UDF10	&	0.28	&	8.34	&	-0.81	&	0.64	&$	0.034	_{	-0.017	}^{+	0.018	}$&$	50	_{-	16	}^{+	16	}$&$	39	_{-	22	}^{+	22	}$\\
\textit{SINS\,/\,zC-SINF}	&	Q2343-BX389	&	2.17	&	10.61	&	1.76	&	0.74	&$	-0.048	_{	-0.030	}^{+	0.032	}$&$	299	_{-	21	}^{+	40	}$&$	56	_{-	15	}^{+	13	}$\\
\textit{SINS\,/\,zC-SINF}	&	Q2346-BX482	&	2.26	&	10.26	&	1.53	&	0.72	&$	0.007	_{	-0.034	}^{+	0.039	}$&$	287	_{-	30	}^{+	63	}$&$	58	_{-	15	}^{+	14	}$\\
\textit{SINS\,/\,zC-SINF}	&	Q2343-BX513	&	2.11	&	10.43	&	1.23	&	0.31	&$	-0.021	_{	-0.075	}^{+	0.075	}$&$	102	_{-	26	}^{+	64	}$&$	55	_{-	28	}^{+	24	}$\\
\textit{SINS\,/\,zC-SINF}	&	Q1623-BX599	&	2.33	&	10.75	&	1.73	&	0.29	&$	-0.036	_{	-0.042	}^{+	0.037	}$&$	139	_{-	36	}^{+	62	}$&$	71	_{-	27	}^{+	18	}$\\
\textit{SINS\,/\,zC-SINF}	&	Q2343-BX610	&	2.21	&	11.00	&	1.56	&	0.53	&$	-0.066	_{	-0.020	}^{+	0.023	}$&$	241	_{-	38	}^{+	62	}$&$	64	_{-	24	}^{+	17	}$\\
\textit{SINS\,/\,zC-SINF}	&	Deep3a-15504	&	2.38	&	11.04	&	1.73	&	0.72	&$	-0.026	_{	-0.013	}^{+	0.013	}$&$	305	_{-	80	}^{+	138	}$&$	63	_{-	15	}^{+	13	}$\\
\textit{SINS\,/\,zC-SINF}	&	Deep3a-6004	&	2.39	&	11.50	&	1.92	&	0.61	&$	-0.032	_{	-0.028	}^{+	0.023	}$&$	362	_{-	126	}^{+	109	}$&$	55	_{-	17	}^{+	11	}$\\
\textit{SINS\,/\,zC-SINF}	&	Deep3a-6397	&	1.51	&	11.08	&	1.63	&	0.69	&$	-0.041	_{	-0.015	}^{+	0.011	}$&$	351	_{-	107	}^{+	138	}$&$	59	_{-	17	}^{+	13	}$\\
\textit{SINS\,/\,zC-SINF}	&	ZC400528	&	2.39	&	11.04	&	1.69	&	0.29	&$	-0.009	_{	-0.049	}^{+	0.045	}$&$	341	_{-	89	}^{+	184	}$&$	28	_{-	15	}^{+	23	}$\\
\textit{SINS\,/\,zC-SINF}	&	ZC400569N	&	2.24	&	11.11	&	1.59	&	0.85	&$	-0.064	_{	-0.014	}^{+	0.016	}$&$	364	_{-	64	}^{+	138	}$&$	43	_{-	21	}^{+	16	}$\\
\textit{SINS\,/\,zC-SINF}	&	ZC400569	&	2.24	&	11.21	&	1.63	&	0.88	&$	-0.050	_{	-0.014	}^{+	0.013	}$&$	312	_{-	13	}^{+	19	}$&$	41	_{-	21	}^{+	23	}$\\
\textit{SINS\,/\,zC-SINF}	&	ZC403741	&	1.45	&	10.65	&	1.33	&	0.26	&$	-0.123	_{	-0.081	}^{+	0.065	}$&$	189	_{-	36	}^{+	73	}$&$	36	_{-	12	}^{+	13	}$\\
\textit{SINS\,/\,zC-SINF}	&	ZC406690	&	2.20	&	10.62	&	1.80	&	0.83	&$	-0.046	_{	-0.027	}^{+	0.028	}$&$	313	_{-	107	}^{+	88	}$&$	60	_{-	15	}^{+	16	}$\\
\textit{SINS\,/\,zC-SINF}	&	ZC407302	&	2.18	&	10.39	&	1.78	&	0.43	&$	-0.023	_{	-0.020	}^{+	0.020	}$&$	217	_{-	40	}^{+	71	}$&$	56	_{-	25	}^{+	11	}$\\
\textit{SINS\,/\,zC-SINF}	&	ZC407376S	&	2.17	&	10.14	&	1.51	&	0.19	&$	-0.021	_{	-0.073	}^{+	0.071	}$&$	89	_{-	45	}^{+	65	}$&$	77	_{-	41	}^{+	37	}$\\
\textit{SINS\,/\,zC-SINF}	&	ZC407376	&	2.17	&	10.40	&	1.65	&	0.65	&$	-0.048	_{	-0.048	}^{+	0.044	}$&$	86	_{-	20	}^{+	24	}$&$	56	_{-	27	}^{+	28	}$\\
\textit{SINS\,/\,zC-SINF}	&	ZC412369	&	2.03	&	10.34	&	1.62	&	0.36	&$	-0.047	_{	-0.051	}^{+	0.050	}$&$	120	_{-	27	}^{+	40	}$&$	75	_{-	27	}^{+	13	}$\\
\hline
\end{longtable}
\twocolumn

\section{Effects of galaxy size}
\label{s:app_galaxysize}
Galaxies evolve significantly in size in the range of redshifts we cover in this work \citep[e.g.,][]{2019ApJ...872L..13M}. Thus, size evolution could potentially explain some of the trends observed in the data that we compile. A common approach to take the galaxy size into account is to express the metallicity gradient as a value normalised to the effective half-light radius, \textit{i.e., } in units of $\rm{dex\,r_e^{-1}}$ \citep[e.g.,][]{2017MNRAS.469..151B}, rather than an absolute gradient measured in $\rm{dex\, kpc}^{-1}$. In this appendix, we study how our results change if we examine normalised rather than absolute gradients. For this purpose, we use the half-light radii reported by the source works we use to compile the observed data (see \autoref{tab:tab2}). The model sits in the natural space of $\rm{kpc}$ as guided by the input self-similar galaxy evolution model \citep{2018MNRAS.477.2716K}, and does not provide an independent estimate of $r_{\rm{e}}$ as a function of galaxy mass. Thus we simply use the mean of the observed half-light radii for galaxies falling in each $v_{\phi}$ bin (see \autoref{fig:grad_sigma}) to obtain the model metallicity gradients in units of $\rm{dex\,r^{-1}_{\rm{e}}}$. We caution that this approach limits our ability to make a fair comparison of the model with the data on metallicity gradients normalised by galaxy size, since we are using the measured effective radii directly in the comparison, rather than predicting them self-consistently.

\begin{figure*}
\includegraphics[width=1.0\columnwidth]{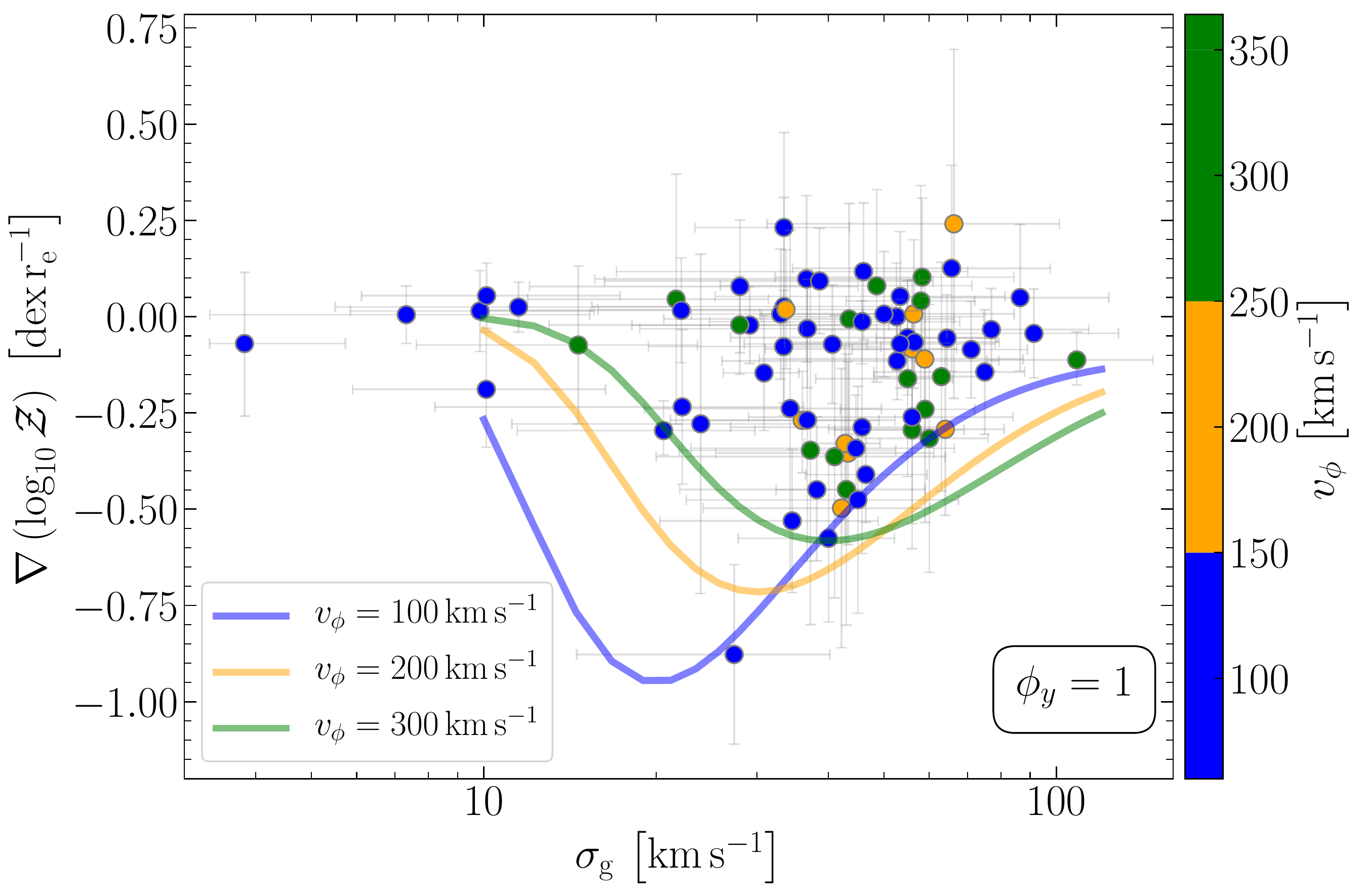}
\includegraphics[width=1.0\columnwidth]{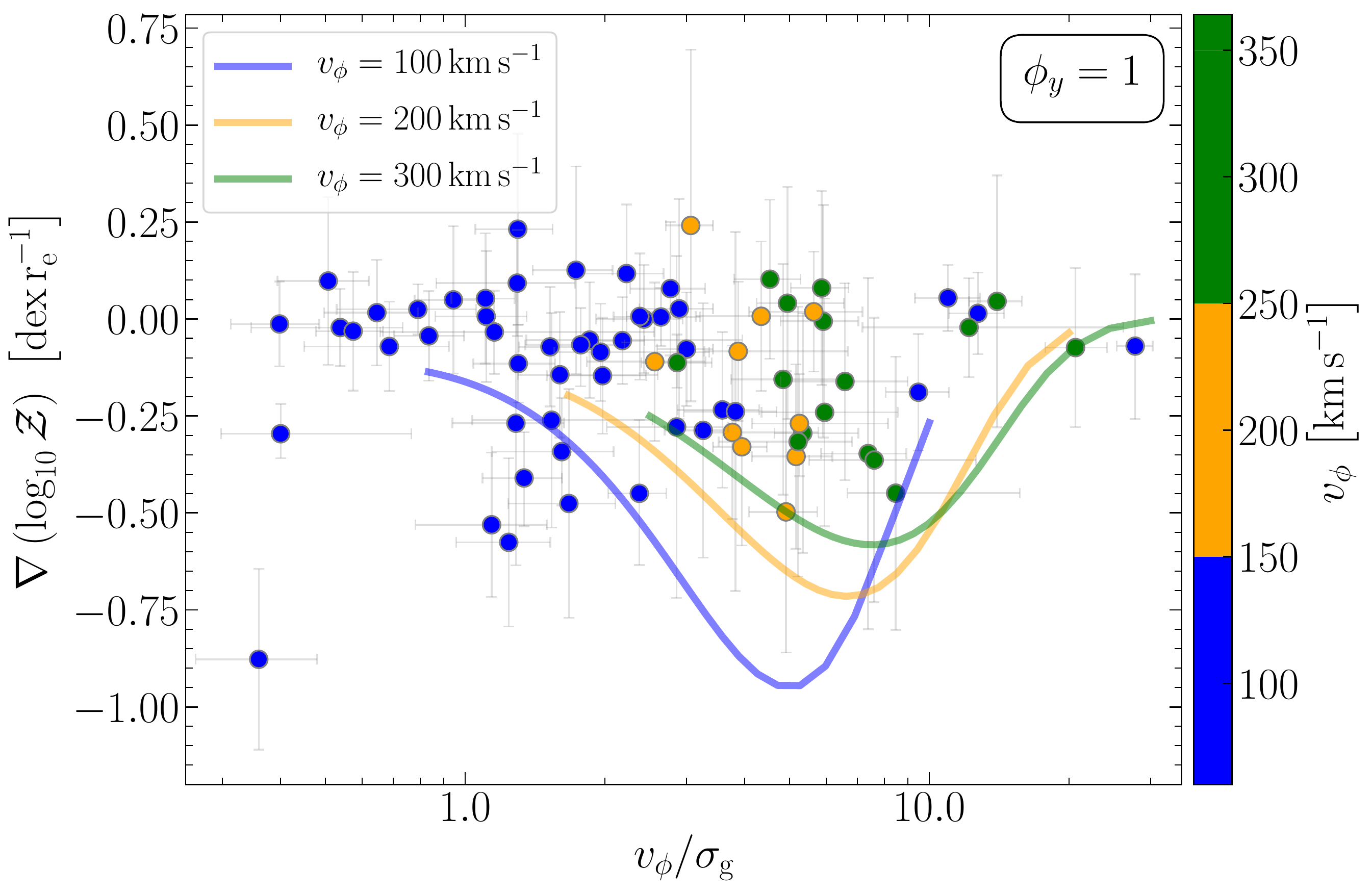}
\caption{\textit{Left panel:} Same as the right panel of \autoref{fig:grad_sigma_model}, but now the metallicity gradients are normalized to the galaxy size and expressed in $\rm{dex\,r_e^{-1}}$. \textit{Right panel:} Same as the right panel of \autoref{fig:grad_vsigma_model}, but with metallicity gradients expressed in $\rm{dex\,r_e^{-1}}$.}
\label{fig:grad_sigma_vsigma_re}
\end{figure*}

\autoref{fig:grad_sigma_vsigma_re} shows the same data and model as in the right panels of \autoref{fig:grad_sigma_model} and \autoref{fig:grad_vsigma_model}, but with the metallicity gradients expressed in the units of $\rm{dex\,r_e^{-1}}$. As expected based on the results in the main text, we find that the results do not significantly change even when the galaxy size is taken into account while studying the trends between metallicity gradients and gas kinematics.

\bsp	
\label{lastpage}
\end{document}